\documentclass[tightenlines,superscriptaddress,floatfixolumn,preprint,footinbib]{revtex4-1} 
\usepackage{graphicx}
\usepackage{amsmath}
\usepackage{amsfonts,amsbsy}
\usepackage{amssymb}

\def\ptasc{p^{\textrm{asc}}_T}
\def\ptrig{p^{\textrm{trig}}_T}
\def\Ntrk{N_{\rm trk}^{\rm offline}}

\def\Np{N_{\rm part}}
\def\as{\alpha_S}
\def\sp{S_\perp}

\def\nc{N_c}
\def\cf{C_F}
\def\qp{ {\bf q}_T } 
 
\def\pp{ {\bf p}_T } 
\def\kp{ {\bf k}_T } 
 
\newcommand{\kpn}[1]{ {\bf k}_{#1\perp} } 
\newcommand{\qpn}[1]{ {\bf q}_{#1\perp} }
\newcommand{\ud}{\mathrm{d}}

\begin{document}

\title{Explanation of systematics of CMS p+Pb high multiplicity di-hadron data at $\sqrt{s}_{\rm NN} = 5.02$ TeV}

\author{Kevin Dusling}
\affiliation{Physics Department, North Carolina State University, Raleigh, NC 27695, USA
}
\author{Raju Venugopalan}
\affiliation{Physics Department, Brookhaven National Laboratory,
  Upton, NY 11973, USA
}

\begin{abstract}
In a recent article (arXiv:1210.3890), we showed that high multiplicity di-hadron proton-proton (p+p) data from the CMS experiment are in excellent agreement with computations in the Color Glass Condensate (CGC) Effective Field Theory (EFT). This agreement of the theory with several hundred data points provides a non-trivial description of both nearside (``ridge") and 
away-side azimuthal collimations of long range rapidity correlations in p+p
collisions.  Our prediction in arXiv:1210.3890 for proton-lead (p+Pb) collisions is consistent 
with results from the recent CMS p+Pb run at $\sqrt{s}_{\rm NN} = 5.02$ TeV for
the largest track multiplicity $N_{\rm track}\sim 40$ we considered. The CMS p+Pb data shows the following striking features: i) a 
strong dependence of the ridge yield on $N_{\rm track}$, with a significantly
larger signal than in p+p for the same $N_{\rm track}$, ii) a stronger $p_T$
dependence than in p+p for large $N_{\rm track}$, and iii) a nearside collimation for large $N_{\rm track}$ comparable to the
awayside for the lower $p_T = p_{T}^{\rm trig.}=p_{T}^{\rm assoc.}$ di-hadron
windows. We show here that these systematic features of the CMS p+Pb di-hadron data are all described by the CGC (with parameters fixed by the p+p data) when we extend our prediction in arXiv:1210.3890 to rarer high multiplicity events. We also predict the azimuthally collimated yield for yet unpublished windows in the $p_{T}^{\rm trig.}$ and $p_{T}^{\rm assoc.}$ matrix. 

\end{abstract}

\maketitle

\section{Introduction}

Rapidity separated di-hadron correlations in high multiplicity events at the LHC
offer sub-femtoscopic scale snapshots of rare configurations constituting the
structure of matter in the colliding hadrons. A largely unexpected discovery at
the LHC by the CMS collaboration in high multiplicity $N_{\rm track} > 110$ proton-proton (p+p) events~\cite{Khachatryan:2010gv} was a collimation in the azimuthal ``nearside" separation ($\Delta \phi\approx 0$) between charged hadrons that have rapidity separations 
$2\leq \left\vert\Delta \eta\right\vert \leq 4$. For recent reviews on this nearside ``ridge" effect, see ~\cite{Li:2012hc,Kovner:2012jm}. In ref.~\cite{Dusling:2012iga}, we showed that this ridge could be explained by ``Glasma graphs"~\cite{Dumitru:2008wn,Dusling:2009ni,Dumitru:2010iy} that arise in the 
Color Glass Condensate (CGC) Effective Field Theory (EFT)~\cite{Gelis:2010nm}.
When the phase space density of gluons in the proton's wavefunction reach
maximal occupancy, or saturation, these graphs are significantly enhanced in
high multiplicity events relative to minimum bias by $\as^{-8}$, a factor of
$10^4$--$10^5$ for typical values of $\as$. This enhancement is a remarkable illustration of
how the power counting changes in different dynamical regions of the EFT. 

Recently, we extended this study significantly~\cite{Dusling:2012cg}, and showed
that a combination of saturation~\cite{Gribov:1984tu,Mueller:1985wy} and BFKL
dynamics~\cite{Balitsky:1978ic,Kuraev:1977fs} in the CGC EFT provides an
excellent description of several hundred data points comprising a matrix (in
uniformly spaced windows in the di-hadron momenta $\ptrig$ and $\ptasc$) of the associated di-hadron yield per trigger versus $\Delta \phi$. A novel feature of this study was the demonstration that BFKL dynamics, which generates gluon emissions between 
the gluons that fragment into triggered hadrons, does an excellent job
describing the awayside spectra. The description is significantly better than
PYTHIA-8~\cite{Khachatryan:2010gv}, and $2\rightarrow 4$ QCD graphs in the
Quasi--Multi--Regge--Kinematics (QMRK), both of which overestimate the awayside
yield, especially at larger momenta. 

In ref.~\cite{Dusling:2012cg}, we also made a prediction for the ridge and the awayside collimation in proton-lead collisions 
at $\sqrt{s}_{NN}=5.02$ TeV at the LHC.  However, as we shall discuss in detail,
the prediction corresponded to a value of $N_{\rm track} \sim 40$ for p+Pb collisions. The magnitude of the signal is comparable to that in p+p collisions at $N_{\rm track} \sim 100$.
Di-hadron data from the p+Pb run at the LHC at  $\sqrt{s}_{NN}=5.02$ TeV are now
available~\cite{CMS:2012qk} and results are available for multiplicities much
larger than than those considered in \cite{Dusling:2012cg}.  These data show the following remarkable features.
i) They exhibit a strong dependence on the number of charged hadron tracks\footnote{Here onwards we will use the CMS notation $\Ntrk$ to discuss the number of charged hadron tracks. Our results will be normalized to the same by equating our multiplicities to their values for the same in minimum bias proton-proton collisions. This point is discussed further on in the text.}, labeled $\Ntrk$ by the CMS collaboration. In particular, it is observed that the
associated di-hadron yield per trigger in p+Pb is significantly larger than the
same signal at the same value of $\Ntrk$ in p+p collisions. ii) Secondly, they
observe a distinct $p_T$ ($\ptrig \sim \ptasc$) dependence of the collimated yield
which is peaked around the same values of $p_T$ in both p+p and p+Pb collisions, but drops off much faster in p+Pb with increasing $p_T$. 
iii) Finally, the di-hadron yield, as a function of $\Delta \phi$, is nearly as high on the nearside as on the awayside for low values of $p_T$, indicating that the long range in $\Delta \eta$ awayside di-jet signal is suppressed relative to the Glasma graph contribution. 

In this paper, we will show that all these novel features of the p+Pb data can be explained systematically in the Color Glass Condensate framework. The parameters in the computations are fixed to be identical to those in our study of p+p collisions in ref.~\cite{Dusling:2012cg}, with the exception being the values of the scale $Q_0 (y_0)$ (whose meaning we shall discuss further) in the proton and lead nuclei.  These are varied at an initial rapidity $y_0$ to take into account the different geometry of lead nuclei relative to that of the projectile proton.  If the systematics of the signal were not reproduced by varying the $Q_0$'s of proton and lead nuclei, there is little freedom left in the framework to vary something else to obtain it.  

The paper is organized as follows. In the next section, we will present the formulae used in the computation of Glasma and BFKL graphs. 
Since all details have been discussed previously in~\cite{Dusling:2012cg} and references therein, we will reintroduce them briefly only for completeness, our focus here being the understanding of the systematics of the new CMS p+Pb data. In section 3,  we will discuss in detail results in the CGC, compare these to the data, and make predictions for as yet unpublished data. In the final section, we will summarize our conclusions, discuss alternative interpretations and further refinements and tests of the CGC framework. 

\section{Glasma and BFKL contributions in the CGC EFT}

\begin{figure}
\centering
\includegraphics[scale=1]{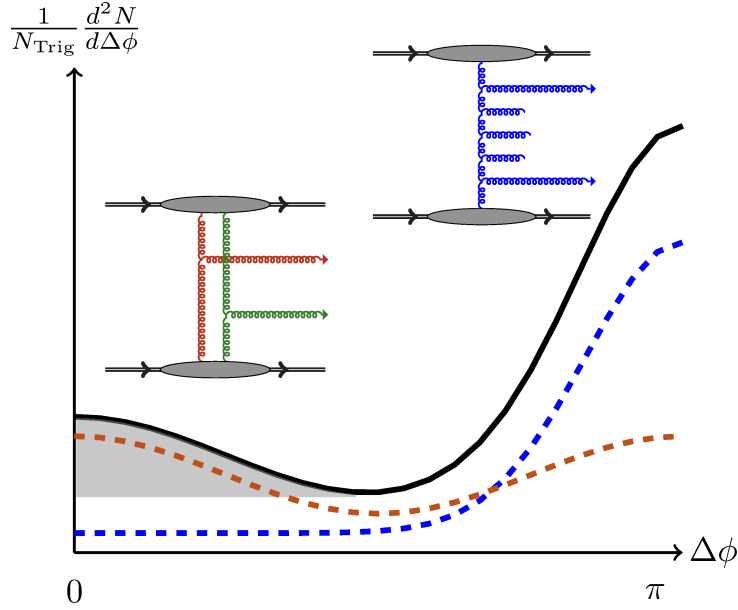}
\caption{Anatomy of di-hadron correlations.  The glasma graph on the left illustrates its 
its schematic contribution to the double inclusive cross-section (dashed orange curve). On the right is the 
back-to-back graph and the shape of its yield (dashed blue curve). The grey blobs denote emissions all the way from beam rapidities to 
those of the triggered gluons. The solid black curve represents the sum of contributions from glasma and back-to-back graphs. The shaded region represents the Associated Yield (AY) calculated using the zero-yield-at-minimum (ZYAM) procedure. Figure from ref.~\cite{Dusling:2012cg}. }
\label{fig:graph}
\end{figure}

The collimated correlated two-gluon production Glasma and BFKL graphs are illustrated in Fig.~(\ref{fig:graph}).  The collimated contributions from all the Glasma graphs can be compactly written as 
\begin{align}
\frac{d^2N_{\rm \sl Glasma}^{\rm \sl corr.}}{d^2\pp d^2\qp dy_p dy_q}
&=  \frac{\alpha_S(\pp)\,\alpha_S(\qp)}{4\pi^{10}}\frac{N_C^2}{(N_C^2-1)^3 \,\zeta}\,\frac{S_\perp}{\pp^2\qp^2}K_{\rm glasma}\nonumber\\
&\times \left[\int_{\kp} (D_1 + D_2)+\sum_{j=\pm}\left(A_1(\pp,j\qp) + \frac{1}{2} A_2(\pp,j\qp)\right)\right] \, .
\label{eq:Glasma-corr}
\end{align}
where we have defined
\begin{align}
D_1 &= \Phi_{A_1}^2(y_p,\kp)\Phi_{A_2}(y_p,\pp-\kp)
\left[\Phi_{A_{2}}(y_q,\qp+\kp)+\Phi_{A_{2}}(y_q,\qp-\kp)\right]\nonumber\,, \\
D_2 &= \Phi_{A_2}^2(y_q, \kp)\Phi_{A_1}(y_p,\pp-\kp)
\left[\Phi_{A_{1}}(y_q,\qp+\kp)+\Phi_{A_{1}}(y_q,\qp-\kp)\right]\, .
\end{align}
These four terms, called the ``single diffractive" and ``interference" graphs in \cite{Dumitru:2008wn}, constitute the leading $\pp/Q_S$ behavior.  Also included is the next order correction in $\pp/Qs$ where we have\footnote{As previously in~\cite{Dusling:2012cg}, the delta distribution is smeared as $\delta(\phi_{pq})\to
\frac{1}{\sqrt{ 2\pi \sigma}}e^{-\frac{\phi_{pq}^2}{2\sigma^2}}$, where $\Delta\phi_{p,q}=\phi_p-\phi_q$ and $\sigma=3\textrm{ GeV}/\pp$
is a $\pp$ dependent width on the order of the saturation scale. The associated yield--the integral over the near-side
signal--is insensitive to details of this smearing.}
$A_1 = \delta^2(\pp+\qp) \left[\mathcal{I}_1^2 + \mathcal{I}_2^2 + 2\mathcal{I}_3^2 \right]$, such that 
\begin{align}
\mathcal{I}_1&=\int_{\kpn{1}} \Phi_{A_1}(y_p,\kpn{1})
\Phi_{A_2}(y_q,\pp-\kpn{1}) \frac{\left(\kpn{1}\cdot
\pp-\kpn{1}^2\right)^2}{\kpn{1}^2\left(\pp-\kpn{1}\right)^2}\;,\nonumber\\
\mathcal{I}_2&=\int_{\kpn{1}}  \Phi_{A_1}(y_p,\kpn{1})
\Phi_{A_2}(y_q,\pp-\kpn{1})\frac{\left|\kpn{1}\times\pp\right|^2}{\kpn{1}^2\left(\pp-\kpn{1}\right)^2}\;,\nonumber\\
\mathcal{I}_3&=\int_{\kpn{1}}  \Phi_{A_1}(y_p,\kpn{1})
\Phi_{A_2}(y_q,\pp-\kpn{1})\frac{\left(\kpn{1}\cdot
\pp-\kpn{1}^2\right)\left|\kpn{1}\times\pp\right|}{\kpn{1}^2\left(\pp-\kpn{1}\right)^2}\;.\nonumber
\end{align}
The other contribution, $A_2$, in Eq.~(\ref{eq:Glasma-corr}) can be expressed as 
\begin{eqnarray}
A_2 =&& \int_{\kpn{1}}
\Phi_{A_1}(y_p,\kpn{1})\Phi_{A_1}(y_p,\kpn{2})
\Phi_{A_2}(y_q,\pp-\kpn{1})
\Phi_{A_2}(y_q,\qp+\kpn{1})
\nonumber\\
&\times&\frac{\left(\kpn{1}\cdot \pp-\kpn{1}^2\right)\left(\kpn{2}\cdot \pp-
\kpn{2}^2\right)+
\left(\kpn{1}\times\pp\right)\left(\kpn{2}\times\pp\right)}{\kpn{1}^2\left(\pp-
\kpn{1}\right)^2}\nonumber\\
&\times&\frac{\left(\kpn{1}\cdot \qp-\kpn{1}^2\right)\left(\kpn{2}\cdot \qp-
\kpn{2}^2\right)+
\left(\kpn{1}\times\qp\right)\left(\kpn{2}\times\qp\right)}{\kpn{2}^2\left(\qp+
\kpn{1}\right)^2}
\label{eq:double-inclusive-5}
\end{eqnarray}
where $\kpn{2}\equiv \pp-\qp-\kpn{1}$.  The above expressions are the result of including all combinatorial combinations of graphs represented by the Feynman diagram to the left in Fig.~\ref{fig:graph}. The combinatorics
is a result of different ways of averaging over strong color sources between the amplitude and complex conjugate amplitude in both projectile and target.

In these expressions\footnote{Other parameters, which are held fixed in p+p and p+Pb are the transverse overlap area $S_\perp$ and the non-perturbative constant $\zeta=1/6$ that represents the effect of soft multigluon interactions, and is independently constrained by p+p multiplicity distributions~\cite{Tribedy:2010ab,Tribedy:2011aa} and real time classical Yang-Mills computations~\cite{Schenke:2012hg}.}, the only function (besides the one loop running coupling constant $\alpha_S$) is the unintegrated gluon distribution (UGD) per unit transverse area 
\begin{equation}
\Phi_A(y,k_\perp) = {\pi N_C k_\perp^2\over 2\alpha_S}\int_0^\infty dr_\perp r_\perp J_0(k_\perp r_\perp)  [1-{\cal T}_A(y,r_\perp)]^2\, 
\label{eq:unint-gluon}
\end{equation}
where ${\cal T}_A$ is the forward scattering amplitude of a quark-antiquark dipole of transverse 
size $r_\perp$ on the target $A$; it, or equivalently, the UGD, is a universal quantity that can be determined by solving the Balitsky-Kovchegov (BK) 
equation~\cite{Balitsky:1995ub,Kovchegov:1999yj} as a function of the rapidity $y=\log\left(x_0/x\right)$.  The forward scattering amplitude ${\cal T}_A(y,r_\perp)$ at the initial scale $x=x_0$ is a dimensionless function of  $r_\perp^2 Q_0^2$, where $Q_0$ is a non-perturbative scale at the initial rapidity. The saturation scale $Q_S$, defined as the transverse momentum defining the peak value of $\Phi$ on the l.h.s of eq.~(\ref{eq:unint-gluon}), is typically a larger scale even at the initial rapidity, and grows rapidly via the BK renormalization group equation with rapidity.  In the BK equation, different impact parameters in the proton/nuclear target are modeled by varying $Q_0$. The minimum-bias (median impact parameter) 
value we choose for the proton $Q_0^2 = 0.168$ GeV$^2$ (corresponding to a
$Q_S^2 \approx 0.7~{\rm GeV}^2$ in the adjoint representation {\it at the initial rapidity}), is the value that gives a best fit to deeply inelastic electron-proton scattering data from HERA~\cite{Albacete:2010sy}. 

We now turn to the double inclusive distribution from the back-to-back BFKL 
graphs shown in Fig.~\ref{fig:graph}. The double inclusive multiplicity can be expressed as~\cite{Colferai:2010wu,Fadin:1996zv}
\begin{eqnarray}
\label{eq:BFKL}
\frac{d^2N_{\rm \sl BFKL}^{\rm \sl corr.}}{d^2\pp d^2\qp dy_p dy_q} &=& \frac{32\,\nc\,
\alpha_s(\pp)\,\alpha_s(\qp)}{ (2\pi)^8 \,\cf}\,\frac{\sp}{\pp^2\qp^2}K_{\rm bfkl}\\
&\times&\int_{\kpn{0}} \int_{\kpn{3}}
\Phi_A(x_1,\kpn{0})\Phi_B(x_2,\kpn{3})\,\mathcal{G}(\kpn{0}-\pp,\kpn{3}+\qp,y_p-y_q)\nonumber
\end{eqnarray}
where $\mathcal{G}$ is the BFKL Green's function
\begin{eqnarray}
\mathcal{G}(\qpn{a},\qpn{b},\Delta y)=\frac{1}{(2\pi)^2}\frac{1}{(\qpn{a}^2 \qpn{b}^2)^{1/2}}\sum_n e^{in\overline{\phi}}\int_{-\infty}^{+\infty} d\nu\textrm{ } e^{\omega(\nu,n)\Delta y}e^{i\nu\ln\left(\qpn{a}^2/\qpn{b}^2\right)}\textrm{   } \, .
\label{eq:BFKL-Green}
\end{eqnarray}
Here $C_F = (\nc^2-1)/2 \nc$, ${\omega(\nu,n)=-2\overline{\alpha}_s\,
\textrm{Re}\left[\Psi\left(\frac{|n|+1}{2}+i\nu\right)-\Psi(1)\right]}$ is the
BFKL eigenvalue, where  $\Psi(z)= d\ln\Gamma(z)/dz$ is the logarithmic
derivative of the Gamma function. Further, we have $\overline{\alpha}_s\equiv
\nc\,\as\left(\sqrt{\qpn{a}\qpn{b}}\right)/\pi$ and 
$\overline{\phi}\equiv \arccos\left(\frac{\qpn{a}\cdot \qpn{b}}{\vert\qpn{a}\vert\textrm{ }\vert \qpn{b}\vert}\right)$. 

\begin{figure}
\includegraphics[width=6in]{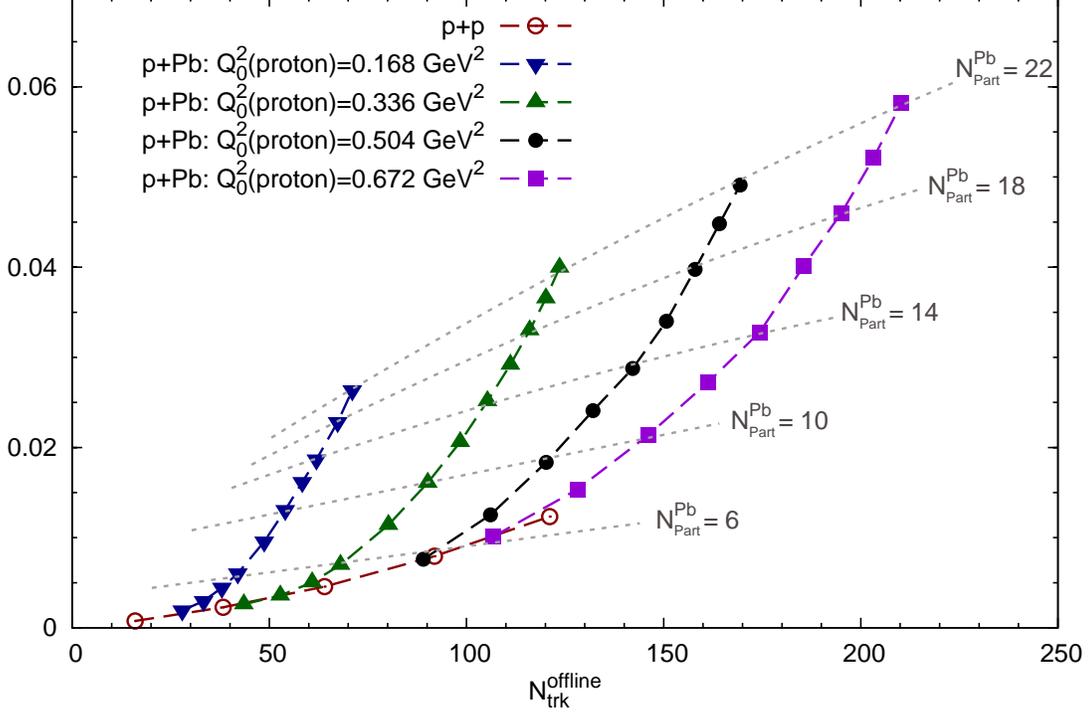}
\caption{The nearside yield per trigger as a function of
$\Ntrk$ for $1\leq p_T \leq 2$, for $p_T=\ptrig=\ptasc$.  Each of the p+Pb curves corresponds
to a fixed initial saturation scale in the proton.  The trajectory of a curve
shows how the yield increases with a larger number of participants in
the nucleus.  The initial saturation scale in the Pb nucleus is related to the
number of participants through $Q_0^2{\rm (lead)} = N_{\rm part}^{\rm Pb}\cdot
0.168~\rm{GeV}^2$.  The values of $Q_0^2{\rm (proton)} = 0.168 - 0.672~{\rm GeV}^2$ (corresponding to saturation scales in the adjoint representation of
$Q_S^2 \approx 0.7 - 1.6~{\rm GeV}^2$) represent estimates these quantities from median (``min.~bias") impact parameters in the proton to very central impact parameters respectively.}
\label{fig:multi}
\end{figure}

As shown in Fig.~(\ref{fig:graph}), Eq.~(\ref{eq:BFKL}) gives a collimated $\Delta \Phi$ contribution exclusively on the away side, peaked at 
$\Delta\Phi = \pi$, while Eq.~(\ref{eq:Glasma-corr}) gives a ``dipole" $\cos(2\Delta\Phi)$-like contributions with maxima at 
$0$ and $\pi$. It's the interplay between these contributions with varying $Q_0$ in projectile and target that describes the systematics of 
the proton-proton and proton-lead data, that we shall now discuss further.

\section{Results}

As noted, all parameters in Eqs.~(\ref{eq:Glasma-corr}) and (\ref{eq:BFKL}) are identical to those describing the proton-proton data. 
To simulate the p+Pb collision, all we do is vary $Q_0^2$ in the proton and lead nuclei. The proton $Q_0^2$ is varied in multiples of 
the ``minimum bias" value of $Q_0^2=0.168$ GeV$^2$ to simulate events that probe more central impact parameters in the proton, where the 
gluon density is considerably higher than the gluon density for the median impact parameter corresponding to minimum-bias events. Likewise, we define the initial saturation scale in lead to be $Q_0^2= N_{\rm part}^{\rm Pb} \cdot
0.168~{\rm GeV}^2$, where $N_{\rm part}^{\rm Pb}$ denotes the number of nucleon participants on the lead side. 

\begin{figure}
\includegraphics[width=3.2in]{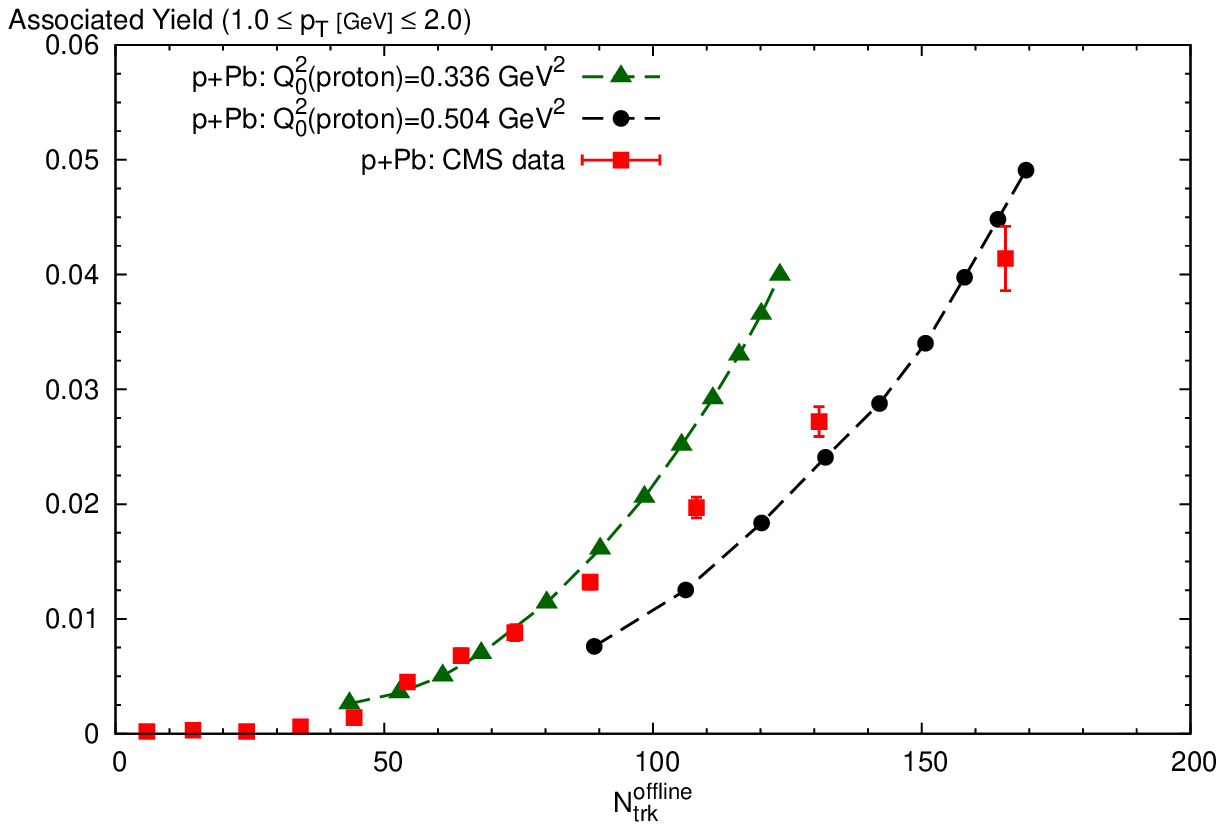}
\includegraphics[width=3.2in]{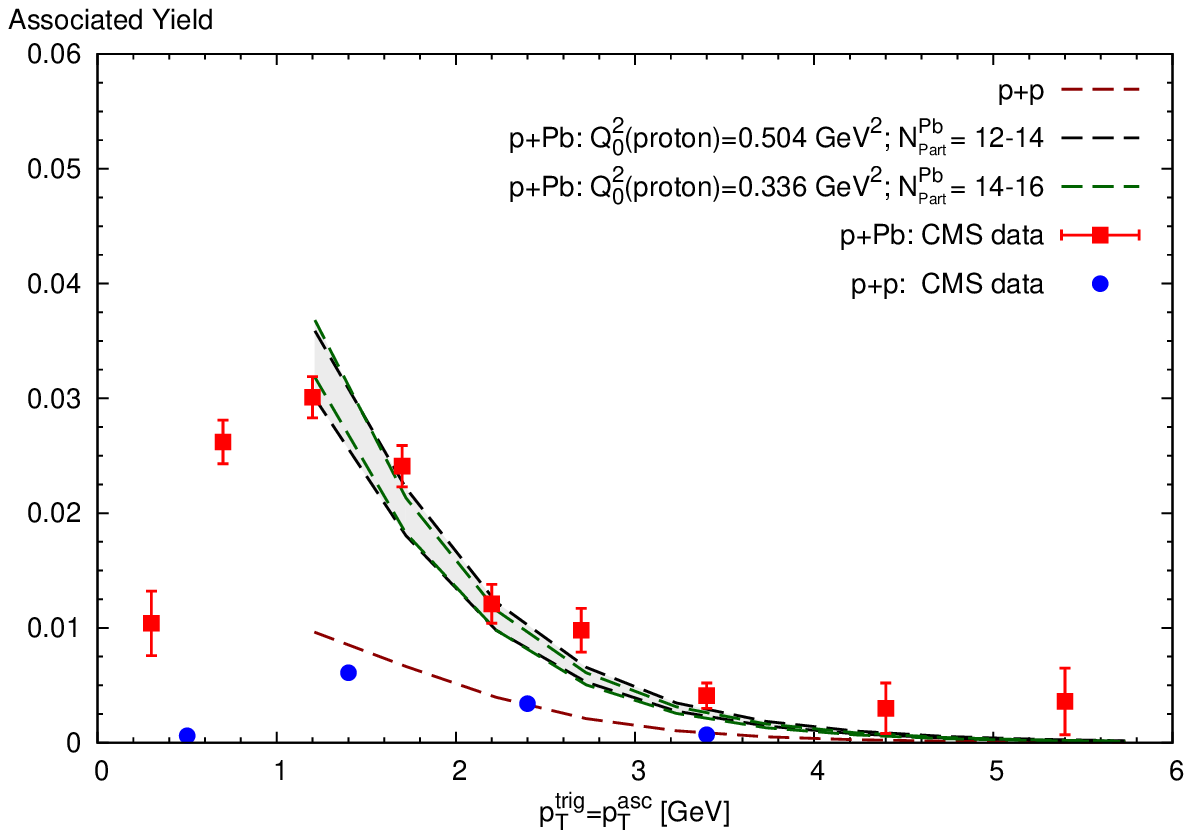}
\caption{Left: The integrated nearside associated yield per trigger as a function of
$N_{\rm trk}^{\rm offline}$ for $1\leq p_T\leq 2$. The two curves on which data from ~\cite{CMS:2012qk} are overlaid are the
$Q_0^2$(proton)=0.336 GeV$^2$ and $Q_0^2$(proton)=0.504 GeV$^2$ results from Fig.~(\ref{fig:multi}).  Right:
The $p_T$ ($\ptrig=\ptasc$) dependence of the associated yield 
for the same $Q_0^2$(proton) values as the previous plot denoted by green
(lower) and black (upper) dashed lines, for two different $N_{\rm part}^{\rm
Pb}$ ranges. The data here are for $N_{\rm trk}^{\rm offline} \geq 110$ that is
approximated (see Fig.~\ref{fig:multi}) by the $N_{\rm part}^{\rm Pb}$ ranges
considered.} 
\label{fig:ay_pPb}
\end{figure}

For our analysis of the CMS data, we define 
\begin{align}
N_{\rm trk}^{\rm offline}=\int_{-2.4-y_{\rm shift}}^{+2.4-y_{\rm shift}}\! \!\!\!\! d\eta \!\!\int_{p_T^{\rm min}}^{p_T^{\rm max}}\!\!\!\! d^2\pp\!\!\int_{0}^1 \!\!  dz \frac{D(z)}{z^2} \frac{dN}{d\eta \,d^2 \pp}\left(\frac{p_{\textrm{T}}}{z}\right)\, .
\label{eq:ntrig}
\end{align}
The single inclusive gluon distribution in the r.h.s is defined as~\cite{Dusling:2012iga}
\begin{align}
\frac{\ud N_1}{\ud y_p\ud^2\pp }
=\frac{\alpha_s N_C}{4\pi^6 (N_C^2-1)}\frac{S_\perp}{\pp^2}
\int_{\kp}\!\!\!
\Phi_{A_1}(y_p,\kp)\,\Phi_{A_2}(y_p,\pp-\kp)\,,
\label{eq:single-incl}
\end{align}
where $y_{\rm shift}=0.465$ is the shift in rapidity in the center-of-mass frame in asymmetrical p+Pb collisions towards the lead fragmentation region. In Eq.~(\ref{eq:ntrig}), the fragmentation functions are chosen,
as in \cite{Dusling:2012cg}, to be the NLO KPP  parametrization~\cite{Kniehl:2000fe} for gluon fragmentation 
to charged hadrons. Finally, in Eq.~(\ref{eq:ntrig}), the transverse overlap area $S_\perp$ is fixed (for $Q_0^2=0.168$ GeV$^2$ in both protons), from minimum bias proton--proton collisions to give $N_{\rm trk}^{\rm offline}=16$, the value quoted by CMS~\cite{Khachatryan:2010gv}.  This value is subsequently held fixed to determine $N_{\rm trk}^{\rm offline}$ as $Q_0^2$ in both the proton and lead nucleus is varied. 

The double inclusive multiplicity is computed as, 
\begin{align}
\label{eq:dihadron}
&\frac{d^2N}{d\Delta \phi} = \int_{-2.4-y_{\rm shift}}^{+2.4-y_{\rm shift}} \!d\eta_p  \,d\eta_q \,\, {\cal A}\left(\eta_p,\eta_q\right) \\
&\!\!\times\int_{p_T^{\rm min}}^{p_T^{\rm max}} \frac{dp_T^2}{2} \int_{q_T^{\rm min}}^{q_T^{\rm max}}\frac{ d q_T^2}{2}\;\int d\phi_p \int d\phi_q\; \delta\left(\phi_p-\phi_q-\Delta\phi\right) \nonumber\\
&\!\!\times \int_{0}^1\!\! dz_1 dz_2 \frac{D(z_1)}{z_1^2}\, \frac{D(z_2)}{z_2^2}
 \frac{d^2N_{}^{\rm \sl corr.}}{d^2\pp d^2\qp d\eta_p d\eta_q}\left(\frac{p_{\textrm{T}}}{z_1},\frac{q_{\textrm{T}}}{z_2},\Delta\phi \right)\nonumber
\end{align}
Bounds on the range of the trigger and associated hadron momenta  are denoted respectively as $p_T^{\rm min (max)}$ and $q_T^{\rm min (max)}$. Likewise, 
$\Delta\eta_{\rm min}(\Delta \eta_{max})=2.0 (4.0)$ denote the pseudo-rapidity gap\footnote{Replacing the rapidity $y$ with the pseudo-rapidity $\eta$  is a good approximation for the $p_T$, $q_T$ of interest.} of hadrons within the experimental acceptance\newline {${\cal A}\left(\eta_p,\eta_q\right)\equiv \theta\left( \vert\eta_p -\eta_q\vert - \Delta\eta_{min}\right)\, \theta\left(\Delta\eta_{\max} - \vert \eta_p-\eta_q\vert\right)$}. 

The associated yield is computed using the Zero-Yield-at-Minimum (ZYAM) procedure,
\begin{equation}
\label{eq:zyam}
\textrm{Assoc. Yield} = \frac{1}{\Ntrk}\int_0^{\Delta\phi_{\rm min.}} \!\!\!\!
d\Delta\phi\left(\frac{d^2N}{d\Delta\phi}-\left.\frac{d^2N}{d\Delta\phi}\right|_{\Delta\phi_{\rm
min}}\right)
\end{equation}
where $\Delta\phi_{\rm min.}$ is the angle at which the two particle correlation strength is minimal. An important point to note is that the transverse overlap area $S_\perp$ cancels out between the numerator and denominator in the r.h.s eliminating a source of uncertainty in di-hadron spectra.  

After these preliminaries, we are now ready to discuss our results. In
Fig.~(\ref{fig:multi}), we plot the integrated associated nearside yield per
trigger (obtained from Eqs.~(\ref{eq:dihadron}) and (\ref{eq:zyam})) versus
$N_{\rm trk}^{\rm offline}$ as determined in Eq.~(\ref{eq:ntrig}). The only
inputs are $Q_0^2$(proton) and $Q_0^2$(lead) = $N_{\rm part}^{\rm Pb}\cdot 0.168$ GeV$^2$. We first point out that the prediction in our
paper~\cite{Dusling:2012cg} for the p+Pb ridge corresponded to $N_{\rm
part}^{\rm Pb}=6$  (which we called ``central") for the left most curve (with
$Q_0^2$(proton) $=0.168$ GeV$^2$). Clearly, this signal is close in magnitude to the high multiplicity p+p ridge signal, if one follows the 
line of sight of the $N_{\rm part}^{\rm Pb}=6$ grey dashed line. This is similar
to the observation in Fig.~3 of~\cite{CMS:2012qk}, where the signal at $N_{\rm trk}^{\rm offline}=60$ in p+Pb is comparable to that in p+p at $N_{\rm trk}^{\rm offline}=100$.

What is particularly striking about Fig.~(\ref{fig:multi}) is the large signal
one obtains as one cranks up both $Q_0^2$(proton) and $N_{\rm part}^{\rm Pb}$
in the lead nucleus. As one goes to larger (rarer) values of $N_{\rm trk}^{\rm
offline}$, one observes that each of the curves grows rapidly. The number of
participants on the lead side is in line with Monte-Carlo Glauber estimates for
not especially rare events~\cite{Bozek:2011if,Albacete:2012xq}.  Interestingly,
rarer events are achieved more efficiently by having gluon distributions at more
central impact parameters in the proton (larger values of $Q_0^2$(proton))
interact than by adding a larger and larger number of participants on the lead side. As is well known, the multiplicity in p+A collisions grows linearly with increasing the saturation scale in the proton, but only logarithmically with the saturation scale in the nucleus~\cite{Dumitru:2001ux}, if the latter is the larger of the two \footnote{We have checked that this scaling is approximately satisfied by our computed single inclusive multiplicity.}. There is also a further effect that when one increases the saturation scale in the nucleus, some of the excess multiplicity is pushed out of the 
detector acceptance, this acceptance having wider coverage in the lead nucleus fragmentation region. 
 In general, however, one will have a multiplicity distribution with $N_{\rm trk}^{\rm offline}$ generated by a number of different impact parameters in the overlap of proton and lead nuclei, and one needs to average over 
the signal on the y-axis of Fig.~(\ref{fig:multi}) with the appropriate weight for a more quantitative analysis. Nevertheless, as seen in the figure, the essential point is that there is no problem obtaining a large associated yield for p+Pb collisions at the LHC for reasonable values of $Q_0^2$ and $N_{\rm part}^{\rm Pb}$. The reasons for this we will discuss at length in the next section.

In Fig.~(\ref{fig:ay_pPb}), we show comparisons of computations of the integrated nearside associated yield per trigger with the CMS data from ~\cite{CMS:2012qk}. In the left plot in this figure, we compare to the data the centrality dependence of the associated yield computed for two different values of $Q_0^2$(proton) while varying $N_{\rm part}$ on the lead side. These curves are the same as the second and third curves (from left) in Fig.~(\ref{fig:multi}), with the same $N_{\rm part}^{\rm Pb}$ values labeling the different points. The $p_T$ distributions (for $\ptrig \simeq \ptasc$) as measured by the CMS collaboration are shown in the right plot of Fig.~(\ref{fig:ay_pPb}).  Also shown is a compilation of four curves from the glasma graph computation obtained by varying the initial saturation scale in the proton from $Q_0^2$(proton) = 0.336 to 0.504 GeV$^2$ for $N_{\rm part}^{\rm Pb}$=14 and 16 or $N_{\rm part}^{\rm Pb}$=12 and 14 respectively.  These configurations were chosen to be representative of the $\Ntrk \geq 110$ centrality class. Clearly, as noted, a given $N_{\rm trk}^{\rm offline}$ can correspond to different combinations of configurations from the proton and lead side. A more realistic computation would include an average over all multiplicities weighted
by the corresponding multiplicity distribution. This caveat aside, we find that
the results in Fig.~(\ref{fig:ay_pPb}) reproduce the $\Ntrk$ and $p_T$ dependence of the associated yield rather well.

\begin{figure}[]
\vspace{50pt}
\includegraphics[width=\textwidth]{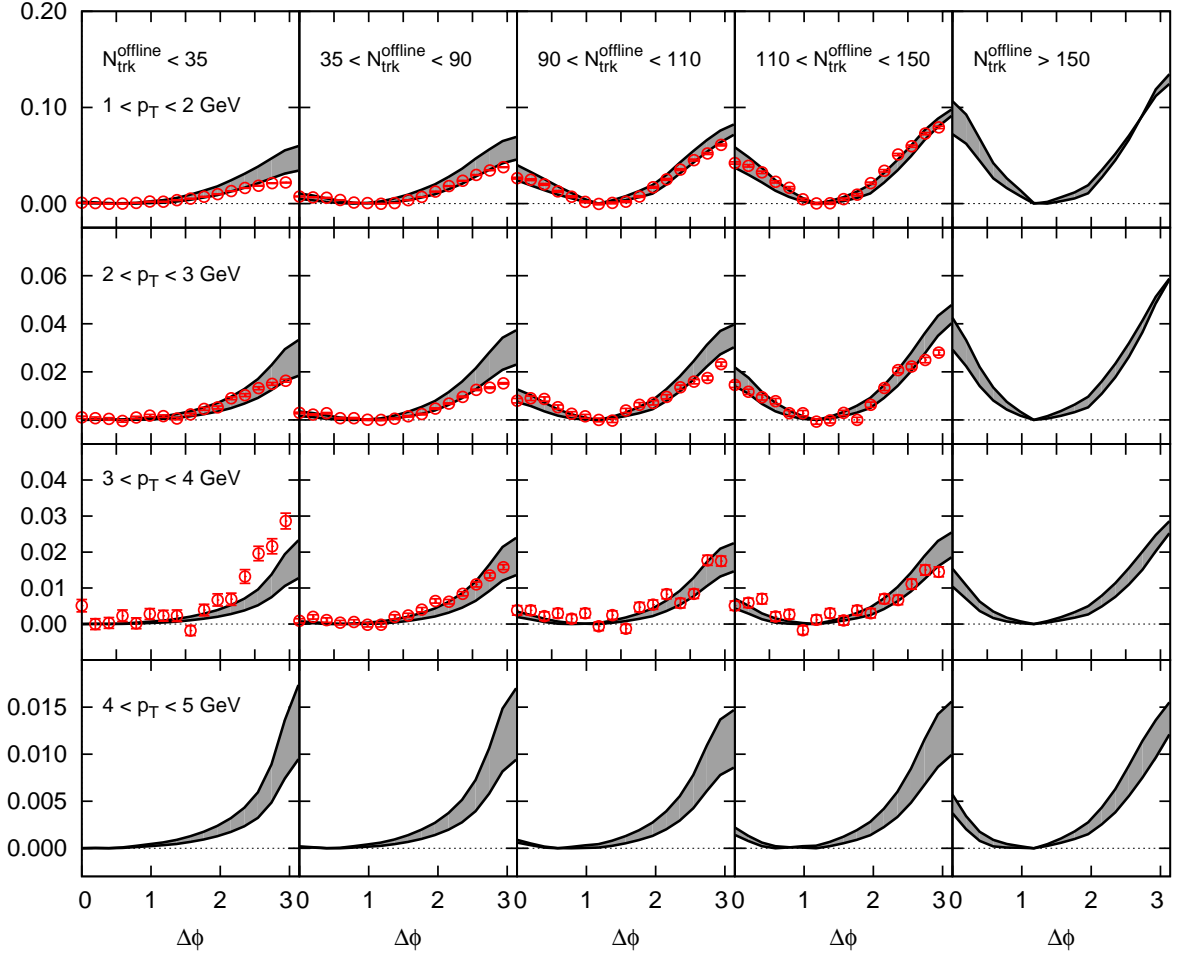}
\caption{Correlated yield $d^2N/d\Delta\phi$ after ZYAM as a function of $\Delta
\phi$ integrated over $2\leq \vert\Delta\eta\vert \leq 4$ for several
multiplicity bins, each for a particular range in $\ptrig=\ptasc$.  The data
points are from the CMS collaboration~\cite{CMS:2012qk}.  The theoretical
curves are the result of adding the glasma and BFKL contributions with the
band representing the variation in results when changing the K-factors from
$K_{\rm glasma}=1, K_{\rm bfkl}=1.1$ to $K_{\rm glasma}=1.3, K_{\rm bfkl}=0.6$.
The results for the different multiplicity windows correspond (from left to right) to: $Q_0^2$(proton)=0.168 GeV$^2$, $N_{\rm part}^{\rm
Pb}=3$; $Q_0^2$(proton)=0.336 GeV$^2$, $N_{\rm part}^{\rm
Pb}=6$; $Q_0^2$(proton)=0.336 GeV$^2$, $N_{\rm part}^{\rm
Pb}=12$; $Q_0^2$(proton)=0.504 GeV$^2$, $N_{\rm part}^{\rm
Pb}=14$; $Q_0^2$(proton)=0.504 GeV$^2$, $N_{\rm part}^{\rm
Pb}=22$. Predictions are shown for very large multiplicity windows and higher values of $\ptrig=\ptasc$. }
\label{fig:pPb_matrix}
\vspace{50pt}
\end{figure}

It is clear at this point that the glasma graphs are able to account for all of
the available systematics of the near-side associated yield.  We now consider a
more differential quantity, the correlated yield as a function of the relative azimuthal
angle $\Delta \phi$ between di-hadrons having momentum $\ptrig$ and $\ptasc$.  In
order to fully understand the $\Delta\phi$ dependence, there are three components
one must have under control as summarized in Fig.~\ref{fig:graph}: firstly, 
the near-side glasma diagrams, which have already been discussed at length.
Secondly, one needs to include the contribution from mini-jet back-to-back graphs in
order to have a quantitative picture of the awayside.  As mentioned here, and as shown clearly 
in our previous work on high multiplicity p+p collisions, di-jet production with BFKL evolution between the triggered
particles is the right framework for assessing this quantity.   Finally, one needs
to have control of the underlying event.  The underlying event is a true
correlation with a distinct $\ptrig$ and $\ptasc$ structure but no angular dependence.  However, we know that
there are other CGC diagrams
\cite{Dumitru:2010mv,Dumitru:2011zz,Kovner:2010xk,Kovner:2011pe} that may
contribute to the underlying event, which do not produce a robust collimation,
hence the importance of the ZYAM procedure to remove these contributions.  We
will briefly discuss some interesting characteristics of the underlying event
within our framework. 

Fig.~(\ref{fig:pPb_matrix}) shows the correlated yield $d^2N/d\Delta\phi$ for
various ranges of $\ptrig$ and $\ptasc$ and centrality classes after performing the ZYAM procedure in
each bin \footnote{Data is also available for $p_T < 1$ GeV; we have not compared the theory predictions to data in this window because the theory systematic errors are large.}.  The shaded band corresponds to one source of uncertainty in our results
from the choice of $K$ factors.  The curve that appears larger on the awayside
correponds to the $K$ factors obtained from our previous analysis of p+p
data ($K_{\rm glasma}=1$ and $K_{\rm bfkl}=1.1$)  with the other curve corresponding to
a new choice (fit by eye) of $K_{\rm glasma}=1.3$ and $K_{\rm bfkl}=0.6$.
The centrality dependence of the result is controlled by our selection of 
representative values of both $Q_0^2$(proton) and $N_{\rm part}^{\rm Pb}$ that
approximately reproduce the mean multiplicity in each centrality class.

\begin{figure}[]
\vspace{50pt}
\includegraphics[width=\textwidth]{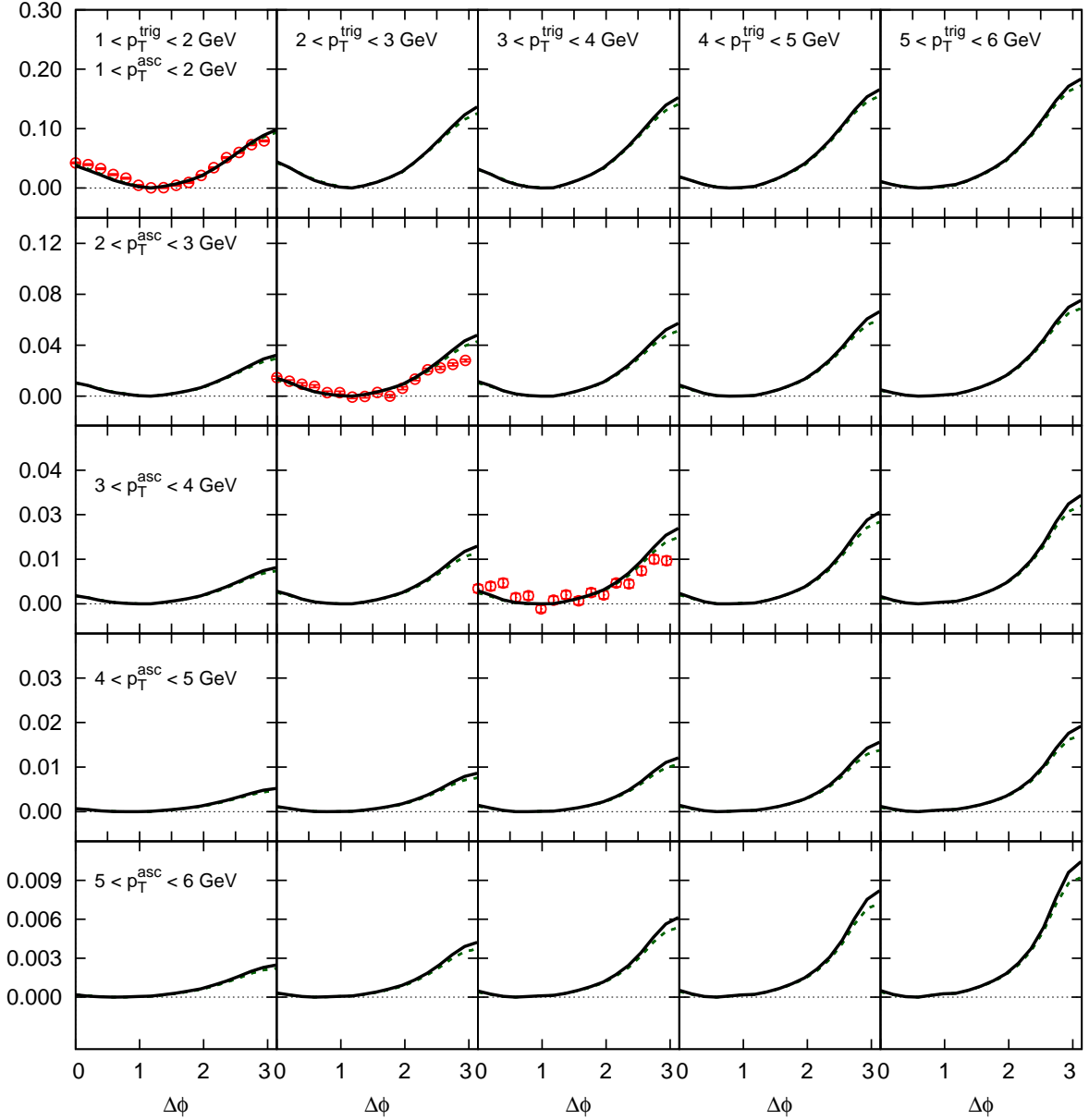}
\caption{Correlated yield $d^2N/d\Delta\phi$ after ZYAM as a function of $\vert \Delta
\phi$ integrated over $2\leq \vert\Delta\eta\vert \leq 4$ for the most
central multiplicity bin $\Ntrk \geq 110$.  The data points are from the
CMS collaboration\cite{CMS:2012qk} and have currently only been provided for the diagonal
components $\ptrig\sim\ptasc$ of the correlation matrix.  The curves
are obtained by adding the glasma contributions ($K=1$) and the BFKL
contribution ($K=1.1$).  The solid black curve is the result for $Q_0^2$(proton) = 0.504 GeV$^2$ on $N_{\rm part}^{\rm Pb}$=14 and the dashed green is
for $Q_0^2$(proton) = 0.336 GeV$^2$ on $N_{\rm part}^{\rm Pb}$=16.}
\label{fig:pPb_matrix_central}
\vspace{50pt}
\end{figure}

In Fig.~(\ref{fig:pPb_matrix_central}), we show results for the highest multiplicity events. CMS p+Pb data is only available at present for three of these windows diagonal in $\ptrig\sim \ptasc$, shown as the red data points.  The curves are
the sum of the glasma contribution and the BFKL
contribution.  The solid black curve is the result for $Q_0^2$(proton)
= 0.504 GeV$^2$ on $N_{\rm part}^{\rm Pb}$=14 and the dashed green is
for $Q_0^2$(proton) = 0.336 GeV$^2$ on $N_{\rm part}^{\rm Pb}$=16.  As before,
we have chosen these values to be representative of the $\Ntrk\geq 110$
centrality class.  A more quantitative result could in principle be obtained by
the appropriate averaging over various events as discussed earlier.  The $K$
factors were chosen to coincide with those extracted from our previous analysis
of p+p collisions ($K_{\rm glasma}=1$ and $K_{\rm bfkl}=1.1$).  There is no reason why these shouldn't be adjusted to p+Pb
collisions \footnote{There is a genuine subtlety in comparison of the glasma
graph $K$-factors in p+p and p+Pb. In Fig.~(2) of \cite{Dusling:2012cg}, the BFKL contribution on the {\it nearside} is ZYAM'ed out and the glasma contribution alone 
with $K=1$ gives a good fit to data. In Fig.~(4), for the matrix, as discussed in the text of \cite{Dusling:2012cg}, the BFKL contribution gives a tiny (relative to awayside) nearside {``anti-collimation"} which is compensated by cranking up the 
glasma $K$ factor to 2.3. However, the BFKL calculation is not reliable on the
nearside and further, it contaminates the tiny but unmistakable glasma signal.
If we make it flat from 0 to $\Delta \phi\sim \pi/2$, all the systematics of p+p
(as in Fig. 2 of \cite{Dusling:2012cg}) would be reproduced with a glasma $K$
factor of unity. In the p+Pb case, because the glasma signal is so large for
high multiplicity windows, the uncertainties in BFKL on the nearside are of not
much import and glasma $K=1$ works quite well.}.

The differential associated yields show that the combination of Glasma and BFKL dynamics provides quite a good 
description of the data without too much fine tuning. The $K$ factors for glasma and BFKL graphs used previously for the high multiplicity p+p results do a good job in many windows of the matrix in Fig.~(\ref{fig:pPb_matrix}) but tend to overpredict the BFKL contribution in some of the windows. {\it Reasonable changes in $K$ values, as shown by the band, are likely to give a finer tuned description of these $\sim 200$ data points along with the additional several hundred data points of p+p data all with $K$ values of order unity. }

As shown in Fig.~(\ref{fig:pPb_matrix_central}), the systematics of the away side signal becoming dominated by the ``dipole"-like Glasma graphs in high multiplicity events is reproduced. It is very non-trivial that the BFKL awayside dynamics is suppressed in 
awayside events such that the combination of Glasma+BFKL on the awayside does not overestimate the awayside signal. This happens because the BFKL di-jet yield per trigger is very weakly dependent on $N_{\rm part}$ since the UGDs in the numerator of Eq.~(\ref{eq:BFKL}) are the same as in the expression for $\Ntrk$ in Eq.~(\ref{eq:ntrig}). 

\begin{figure}
\includegraphics[width=3.2in]{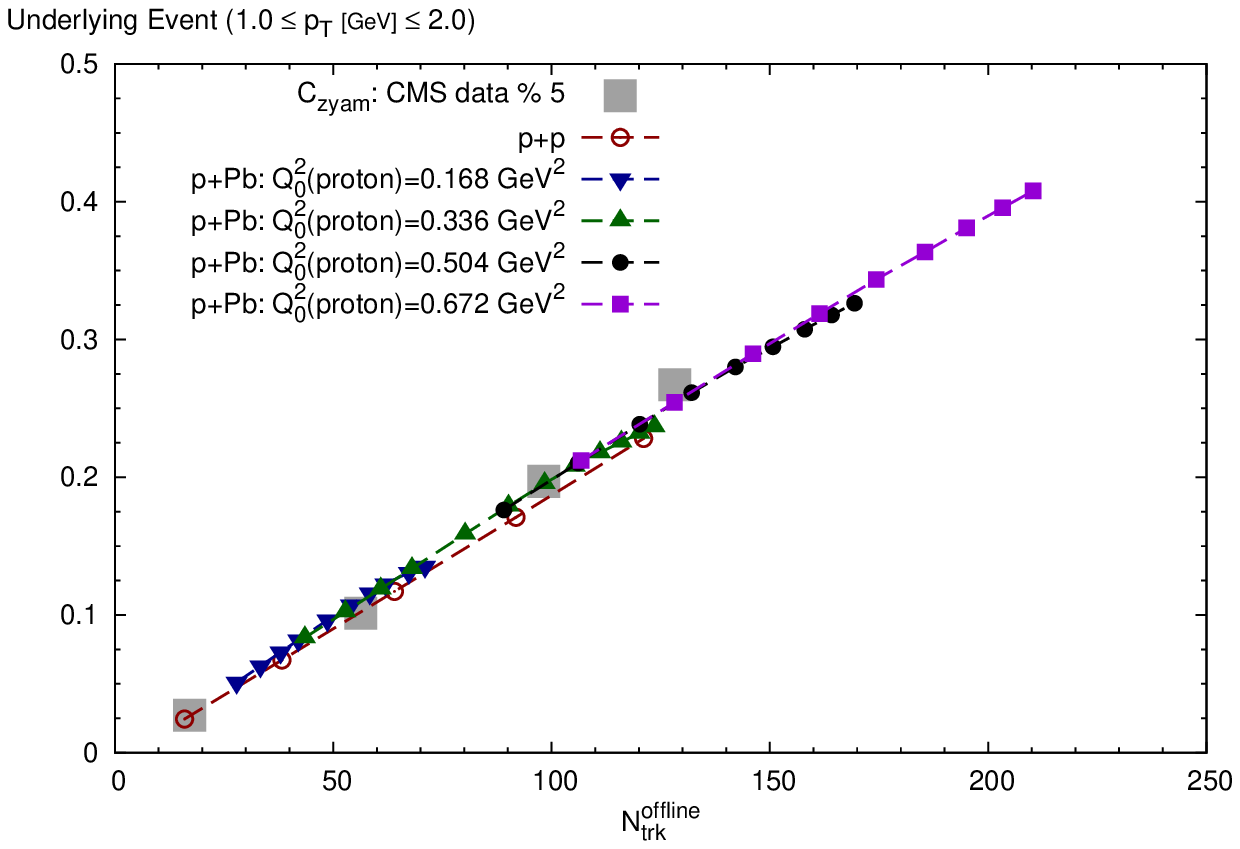}
\includegraphics[width=3.2in]{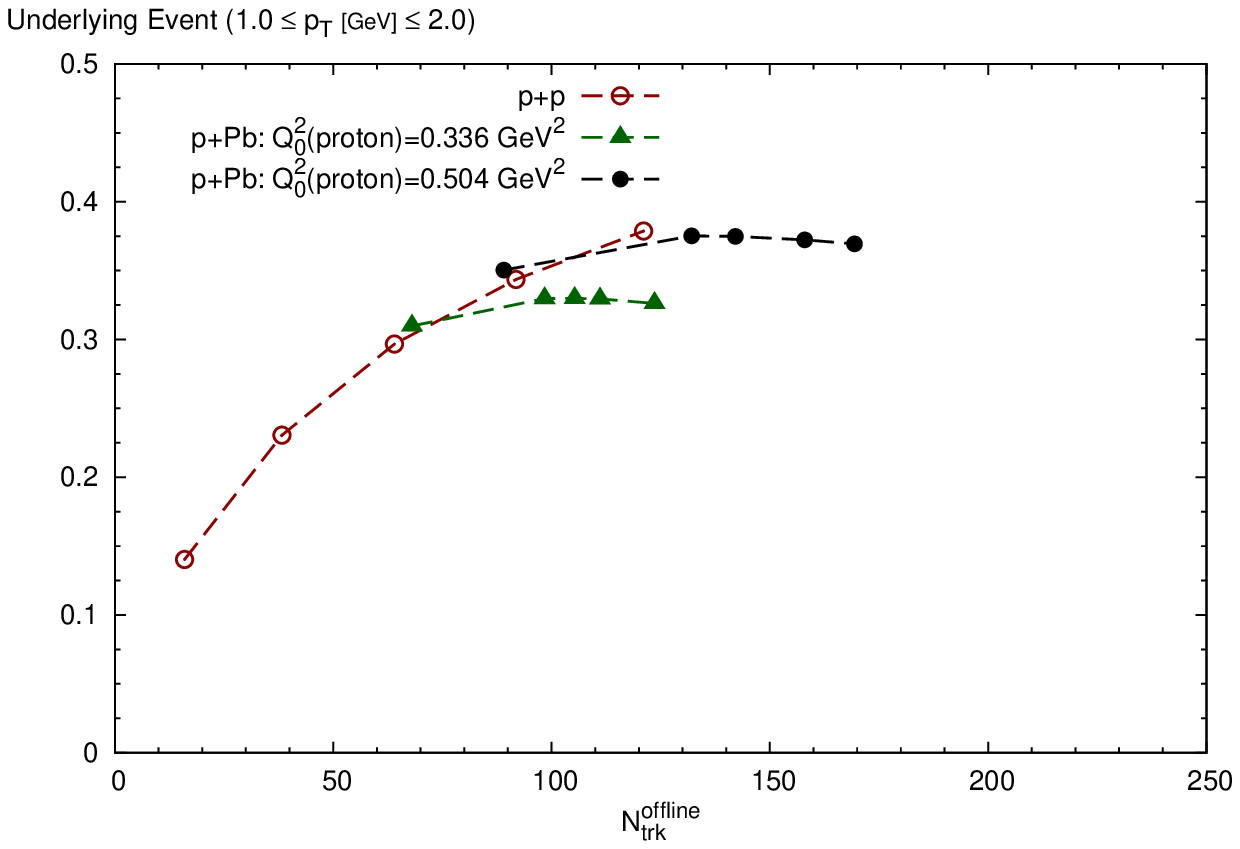}
\caption{Left: The underlying glasma event per trigger (equivalent to $C_{\rm zyam}$) in ~\cite{CMS:2012qk}) as a function of
$\Ntrk$ for $1\leq \ptrig,\ptasc \leq 2$.  The large grey squares are the experimental values of $C_{\rm ZYAM}$ divided by a 
factor 5. All other symbols denote theory computations with a fixed initial scale $Q_0$ in the proton and varying values of 
$N_{\rm part}^{\rm pPb}$.  Right: The underlying BFKL event per trigger for $Q_0^2=0.336$ GeV$^2$ and $Q_0^2=0.504$ GeV$^2$ and varying values of $N_{\rm part}$. See text for discussion.}
\label{fig:ue}
\end{figure}

Finally, it is interesting to examine the $\Ntrk$ dependence of the correlated $\Delta \phi$-independent yield. 
 Fig.~(\ref{fig:ue}) (left) demonstrates that the yield depends on $\Ntrk$ alone for differering values of $Q_0^2$(proton) and $N_{\rm part}$, while the collimated signal shown in Fig.~(\ref{fig:multi}) clearly has a more complex structure. In particular, unlike the case of the collimated yield, the p+p and p+Pb underlying event contributions lie on the same curve. Also shown in Fig.~(\ref{fig:ue}) are the data points for $C_{\rm ZYAM}$ from \cite{CMS:2012qk} divided by a factor of 5. It is very interesting that the
data follow the same $\Ntrk$ scaling as the glasma graphs.  The BFKL contribution to the associated yield is shown in 
Fig.~(\ref{fig:ue}) (right).  It clearly does not have the same $\Ntrk$ scaling, and is approximately of the same magnitude  as the glasma underlying event at $\Ntrk\sim 100$. However, as we noted previously~\cite{Dusling:2012cg}, recent  computations~\cite{Caporale:2012qd,Colferai:2010wu} show that the $\phi$-independent NLLx contributions are a factor of 2-3 {\it below} the LLx contribution \footnote{This may be contrasted to the collimated NLLx contributions that are 10\%-30\% corrections to LLx, and will be significantly accounted for already by the running coupling effects we have included.}.  Hence, there is ample room for other contributions in the CGC that only produce a $\Delta \phi$-independent contribution~\cite{Dumitru:2010mv,Dumitru:2011zz,Kovner:2010xk,Kovner:2011pe}  as long as they give the same scaling as the glasma graphs. 

\section{Discussion and Outlook}

We showed in the previous section that the CGC EFT gives a good description of the novel systematics of proton-lead di-hadron correlations that are long range in rapidity and collimated in the azimuthal angle. This is important because an identical analysis previously gave very good agreement with the CMS data for di-hadron correlations in high multiplicity p+p events. We conclude that the origins of the proton-lead effect are  the same as the one in proton-proton collisions and unlike nucleus-nucleus collisions, where the systematics of the associated yield is dominated by flow~\cite{Dusling:2012iga}. A simple but apt analogy that exemplifies our conclusion is that a bullet shot through a plane of glass has an interaction cross-section closer to the size of the bullet and not that of the glass. 

But what is the deeper origin of the effect ? The systematic features of the
comparison to data are consistent with the following picture of the proton-lead
interaction. As shown previously~\cite{Kowalski:2006hc,Tribedy:2010ab} on the
basis of HERA electron-proton diffractive data, the saturation scale in the
proton has a strong impact parameter dependence which we have modeled here with
different values of $Q_0^2$(proton). In proton-proton collisions, 
$Q_0^2$(proton) = 0.168 GeV$^2$, the value at the median impact parameter is more likely; rare events that correspond to the higher $Q_0^2$ which  produces the high multiplicity collisions (and the ridge) are very unlikely. In contrast, in a proton-lead collision,  any given $\Ntrk$ has a higher probability to be generated  by a larger $Q_0^2$(proton) than the median value. This is because the likelihood that gluons at small impact parameters in the proton interact is much larger when the proton is scattering off many nucleons along its path, as in a lead nucleus. Such events are more likely to dominate the probability $P_N$ for a given $\Ntrk$. That would explain why the values of the associated yield seen in Fig.~(\ref{fig:ay_pPb}) are more compatible with the larger $Q_0^2$(proton) values.  

\begin{figure}
\includegraphics[width=4in]{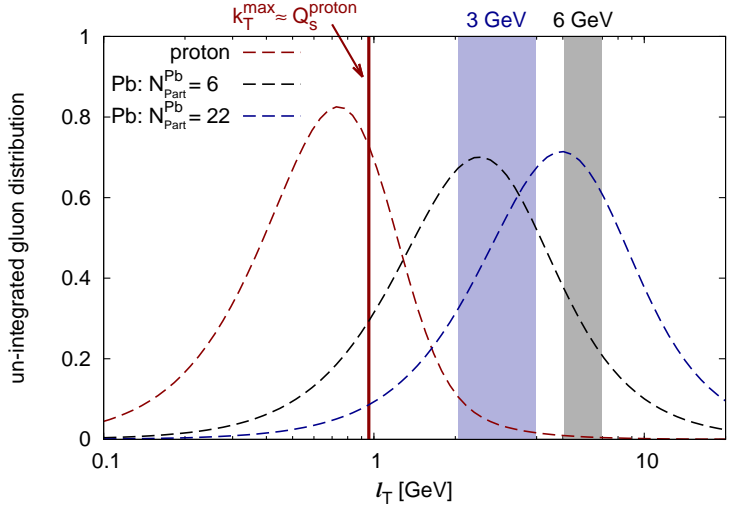}
\caption{The unintegrated gluon distribution (UGD) of a min.~bias proton ($Q_0^2$(proton) = 0.168 GeV$^2$) and for lead nuclei for two different values of $\Np$.
The former is evolved to rapidity Y=2 and the latter are evolved to Y=4.  The red vertical line  $k_T\sim Q_s$(proton) denotes the peak of the integrand in Eq.~(\ref{eq:discuss1}). The remaining angular integral in Eq.~(\ref{eq:discuss1}) is constrained by kinematics to lie within the blue (gray) shaded region for $\vert\pp\vert=3~(6)$~GeV. The collimated signal is determined by the curvature in the shaded regions--flat UGDs are uncollimated.}
\label{fig:wave}
\end{figure}

The prior discussion addresses why larger $Q_0^2$(proton) are more relevant for a given $\Ntrk$. But it does not explain why the associated yield is so large in p+Pb for any $Q_0^2$(proton) as one varies $Q_0^2$(lead) as shown in Fig.~(\ref{fig:multi}). The underlying reason is a subtle form of quantum entanglement. To simplify the discussion, we will consider only one of the Glasma diagrams responsible for the near side collimation evaluated \footnote{For the qualitative arguments in this section, this assumption is sufficient because the glasma signal, to first approximation, is boost invariant.} at $y_p=y_q=0$.  In this case, the two particle correlation is proportional to
\begin{equation}
d^2 N \propto S_\perp \int d^2 k_\perp \Phi_A^2 (\kp) \Phi_B (\vert\pp -\kp\vert) \Phi_B (\vert\qp - \kp\vert) \, .
\label{eq:discuss1}
\end{equation}
To ascertain why the above expression yields a signal that is collimated, namely, a larger signal for $\pp=\qp$ as opposed to the signal when $\pp\neq \qp$, let us consider for simplicity $\vert\pp\vert=\vert\qp\vert$.  This condition, and application of the Cauchy–-Schwarz inequality \footnote{Define $f(\kp)=\Phi_A(\kp)\Phi_B(\protect\vert\pp+\kp\protect\vert)$ and $g(\kp)=\Phi_A(\kp)\Phi_B(\protect\vert\qp+\kp\protect\vert)$. The two dimensional form of the Cauchy–-Schwarz inequality is
\begin{align}
\int d^2k_\perp \;f(\kp)g(\kp)\leq \sqrt{\int d^2\kp\;\left|f(\kp)\right|^2} \sqrt{\int d^2\kp\;\left|g(\kp)\right|^2}\protect\nonumber 
\end{align}
with the equality holding iff $f(\kp)\propto g(\kp)$.} leads to the condition  
\begin{equation}
\int d^2 k_\perp \Phi_A^2 (\kp)\; \Phi_B (\vert\pp -\kp\vert)\; \Phi_B (\vert\qp - \kp\vert) \leq \int d^2 k_\perp \Phi_A^2 (\kp)\; \Phi_B^2 (\vert\pp -\kp\vert) \,.
\end{equation} 
When the equality holds, there is clearly no collimation because the r.h.s does not depend on $\Delta\phi_{pq}$.  However, the equality holds if and only if $\Phi(\vert\pp +\kp\vert) \propto \Phi(\vert\qp +\kp\vert)$, which is only satisfied if the un-integrated gluon distribution is flat within the available phase space.  Fig.~(\ref{fig:wave}) clearly shows that the unintegrated gluon distributions are not flat. Therefore, on 
very general grounds, we expect a collimation from the structure of the two particle correlation in Eq.~(\ref{eq:discuss1}). 

Now that we have argued on very general grounds that there must be a collimation, we would like to understand the scaling of the yield with $\Ntrk$ and $N_{\rm part}$ as seen in Fig.~(\ref{fig:multi}).  As seen in Fig.~(\ref{fig:ue}), the underlying event (which characterizes the overall
normalization of the signal) scales linearly with $\Ntrk$.  Any $\Ntrk$
dependence in Fig.~(\ref{fig:multi}) is therefore a consequence of the $\Ntrk$ scaling of
the normalization.  In addition to this $\Ntrk$ scaling, their is ridge collimation that grows rapidly with $N_{\rm part}$.  We will now discuss in turn both aspects of the systematics of the observed signal. 

From the previous discussion, the underlying event has the form
\begin{equation}
\textrm{UE}\propto
\frac{\int d^2 k_\perp \Phi_A^2 (\kp) \Phi_B^2 (\vert\pp -\kp\vert)}{\int d^2
k_\perp \Phi_A (\kp) \Phi_B (\vert\pp -\kp\vert)}\;,
 \label{eq:discuss2}
\end{equation}
where the term in the denominator is itself proportional to $\Ntrk$.  Because the $\Phi$'s, as shown in Fig.~(\ref{fig:wave}), are bell shaped
curves peaked about the saturation scales, one can deduce by inspection \footnote{This is confirmed by a simple analysis treating the 
$\Phi$'s as Gaussian wavefunctions.} that the numerator scales as $\left(\Ntrk\right)^2$, and hence the ${\rm UE}\propto \Ntrk$.  

We now address the additional $\Np$ scaling that is observed only in the collimated 
associated yield (CY).  This can be characterized by looking at
that ratio of the signal from Eq.~(\ref{eq:discuss1}) evaluated at \footnote{In addition to Eq.~(\ref{eq:discuss1}), there is a contribution that 
has a maximum at $\pi$ and a minimum at $0$. Its the sum of the two contributions that gives the {``dipole"}-like signal with a minimum at 
$\pi/2$.} $\Delta\phi_{pq}=0$ to that at $\Delta\phi_{pq}=\pi$ for
$\vert\pp\vert=\vert\qp\vert$, 
\begin{equation}
\textrm{CY}\propto 
\frac{\int d^2 k_\perp \Phi_A^2 (\kp) \Phi_B^2 (\vert\pp -\kp\vert)}{\int d^2
k_\perp \Phi_A^2 (\kp) \Phi_B (\vert\pp -\kp\vert)\Phi_B (\vert\pp +\kp\vert)}\;.
 \label{eq:discuss3}
\end{equation}
The $\Ntrk$ scaling cancels in this ratio. To see simply how the additional collimation arises, consider the 
extreme scenario where the $\Phi$s are peaked strongly enough to be 
considered Dirac delta distributions.  Working within this approximation we
can easily perform the integral over $d^2 k_\perp$, whereby 
$\Phi_A^2 (\kp)$ fixes $\vert\kp\vert=Q_A$ and the angular integral over ${\Phi_B(\vert\pp -\kp\vert)}$ fixes 
\begin{equation}
\phi=\arccos\left(\frac{Q_B^2-Q_A^2-p_T^2}{2 p_T Q_A}\right)\,.
\end{equation}
After making these substitutions \footnote{Jacobian factors cancel between numerator and denominator.}, we are left with
\begin{equation}
\textrm{CY}\propto 
\frac{\Phi_B (Q_B)}{\Phi_B\left(\sqrt{2p_T^2+2Q_A^2-Q_B^2}\right) }
 \label{eq:discuss4}
\end{equation}
The collimated signal is always larger than unity \footnote{The square root in the denominator implies finite support for $Q_B$ in the narrow 
range $1 \leq Q_B/\sqrt{2p_T^2+2Q_A^2} \leq \sqrt{2}$, which is an artifact of taking Delta
function approximation for the wave-functions.} since the maximum of $\Phi_B$ is at $Q_B$.  As $Q_B$ is increased (while keeping $p_T$ and
$Q_A$ fixed) the wave function in the denominator is probed further away from its
maximum leading to a larger collimation. If we make a Gaussian approximation for the remaining wavefunctions in Eq.~(\ref{eq:discuss4}), we find a rapid growth in the collimated signal with $\Np$.  For the region where $Q_B\gtrsim Q_A$, we find
\begin{equation}
\textrm{CY} \propto 1+\frac{1}{Q_A^2}\left(Q_B-Q_A\right)^2\,,
\end{equation} 
which grows as $\sim \Np$. 

To summarize the discussion, the behavior of the associated yield is a consequence of the quantum entanglement of the wavefunctions of correlated gluons in both the projectile and the target. Since two gluons from both projectile and target participate, one obtains the overlap of 
four wavefunctions. Besides energy-momentum constraints on the wavefunctions, the signal is sensitive to the detailed structure of these wavefunctions.  This includes both the density of gluons with varying impact parameter, as well as the $p_T$ dependence of the gluon distributions for a fixed impact parameter.  With the stated simple yet fairly general assumptions, the scaling of the collimated yield and the underlying event as a function of $\Ntrk$ and $\Np$ is reproduced. The physics of saturation is absolutely crucial: firstly, on a ``global" level because the glasma graphs are tremendously enhanced due to the large phase space occupancy of gluons, but equally so  because the observed signal is sensitive to detailed features of the CGC EFT. 

The theoretical framework employed here can be further improved. An important step is to self-consistently include multiple scattering effects alongside the rapidity evolution of two gluon production. The framework to do this has been developed but not implemented numerically yet~\cite{Gelis:2008sz}. Another improvement is to quantify the NLLx contributions to the collimated yield and the underlying event for the kinematics of interest~\cite{Caporale:2012qd,Colferai:2010wu}. Not least, the contributions of leading $N_c$ multi-gluon correlators and possible pomeron loop effects need to be quantified~\cite{Kovner:2012jm}. 

Finally, while this work was in preparation, a preprint appeared which interprets the effect as due to hydrodynamic flow~\cite{Bozek:2012gr}. As noted previously, we believe the ridge in p+Pb collisions to be more analogous to high multiplicity p+p collisions than Pb+Pb collisions: the systematics of the study here lends weight to this conclusion. For the p+p case, we showed that inclusion of flow~\cite{Dusling:2012iga} changes the structure of the associated yield from that observed in the data even for modest flow velocities.  While some multiple scattering cannot be categorically ruled out, a consistent hydrodynamic description is challenging for systems with transverse sizes the order of the proton size because of poor convergence of the gradient expansion and the short lifetime of the system.  It will be interesting to see whether the hydrodynamic description of ~\cite{Bozek:2012gr} holds for a wider $p_T$ and centrality range than shown. In this regard, it is important that one include the non-flow jet-like BFKL contribution that provides a significant contribution to the awayside yield.

\section*{Acknowledgements}
 K.D.  and  R.V are  supported by the US Department of Energy under DOE Contract Nos.
DE-FG02-03ER41260 and DE-AC02-98CH10886 respectively.  This research used resources of the National Energy Research Scientific Computing Center, which is supported by the Office of Science of the U.S.  Department of Energy under Contract No. DE-AC02-05CH11231.  We would like to acknowledge useful conversations with Adam Bzdak, Adrian Dumitru, Alex Kovner, Wei Li and Larry McLerran, and thank Wei Li for sending us the tabulated CMS p+Pb data.


\begin{thebibliography}{10}%
\makeatletter
\providecommand \@ifxundefined [1]{%
 \ifx #1\undefined \expandafter \@firstoftwo
 \else \expandafter \@secondoftwo
\fi
}%
\providecommand \@ifnum [1]{%
 \ifnum #1\expandafter \@firstoftwo
 \else \expandafter \@secondoftwo
\fi
}%
\providecommand \enquote [1]{``#1''}%
\providecommand \bibnamefont  [1]{#1}%
\providecommand \bibfnamefont [1]{#1}%
\providecommand \citenamefont [1]{#1}%
\providecommand\href[0]{\@sanitize\@href}%
\providecommand\@href[1]{\endgroup\@@startlink{#1}\endgroup\@@href}%
\providecommand\@@href[1]{#1\@@endlink}%
\providecommand \@sanitize [0]{\begingroup\catcode`\&12\catcode`\#12\relax}%
\@ifxundefined \pdfoutput {\@firstoftwo}{%
 \@ifnum{\z@=\pdfoutput}{\@firstoftwo}{\@secondoftwo}%
}{%
 \providecommand\@@startlink[1]{\leavevmode}%
 \providecommand\@@endlink[0]{}%
}{%
 \providecommand\@@startlink[1]{%
  \leavevmode
  \pdfstartlink
   attr{/Border[0 0 1 ]/H/I/C[0 1 1]}%
   user{/Subtype/Link/A<</Type/Action/S/URI/URI(#1)>>}%
  \relax
 }%
 \providecommand\@@endlink[0]{\pdfendlink}%
}%
\providecommand \url  [0]{\begingroup\@sanitize \@url }%
\providecommand \@url [1]{\endgroup\@href {#1}{\urlprefix}}%
\providecommand \urlprefix [0]{URL }%
\providecommand \Eprint[0]{\href }%
\@ifxundefined \urlstyle {%
  \providecommand \doi [1]{doi:\discretionary{}{}{}#1}%
}{%
  \providecommand \doi [0]{doi:\discretionary{}{}{}\begingroup
  \urlstyle{rm}\Url }%
}%
\providecommand \doibase [0]{http://dx.doi.org/}%
\providecommand \Doi[1]{\href{\doibase#1}}%
\providecommand \bibAnnote [3]{%
  \BibitemShut{#1}%
  \begin{quotation}\noindent
    \textsc{Key:}\ #2\\\textsc{Annotation:}\ #3%
  \end{quotation}%
}%
\providecommand \bibAnnoteFile [2]{%
  \IfFileExists{#2}{\bibAnnote {#1} {#2} {\input{#2}}}{}%
}%
\providecommand \typeout [0]{\immediate \write \m@ne }%
\providecommand \selectlanguage [0]{\@gobble}%
\providecommand \bibinfo [0]{\@secondoftwo}%
\providecommand \bibfield [0]{\@secondoftwo}%
\providecommand \translation [1]{[#1]}%
\providecommand \BibitemOpen[0]{}%
\providecommand \bibitemStop [0]{}%
\providecommand \bibitemNoStop [0]{.\EOS\space}%
\providecommand \EOS [0]{\spacefactor3000\relax}%
\providecommand \BibitemShut [1]{\csname bibitem#1\endcsname}%
\bibitem{Khachatryan:2010gv}%
  \BibitemOpen
  \bibfield{author}{%
  \bibinfo {author} {\bibfnamefont{V.}~\bibnamefont{Khachatryan}} \emph{et~al.}
  (\bibinfo {collaboration} {CMS Collaboration}),\ }%
  \bibfield{journal}{%
  \Doi{10.1007/JHEP09(2010)091}{\bibinfo {journal} {JHEP}}\ }%
  \textbf{\bibinfo {volume} {1009}},\ \bibinfo {pages} {091} (\bibinfo {year}
  {2010}),\ \Eprint{http://arxiv.org/abs/1009.4122}{arXiv:1009.4122 [hep-ex]}%
  \bibAnnoteFile{NoStop}{Khachatryan:2010gv}%
\bibitem{Li:2012hc}%
  \BibitemOpen
  \bibfield{author}{%
  \bibinfo {author} {\bibfnamefont{W.}~\bibnamefont{Li}},\ }%
  \bibfield{journal}{%
  \Doi{10.1142/S0217732312300182}{\bibinfo {journal} {Mod.Phys.Lett.}}\ }%
  \textbf{\bibinfo {volume} {A27}},\ \bibinfo {pages} {1230018} (\bibinfo
  {year} {2012}),\ \Eprint{http://arxiv.org/abs/1206.0148}{arXiv:1206.0148
  [nucl-ex]}%
  \bibAnnoteFile{NoStop}{Li:2012hc}%
\bibitem{Kovner:2012jm}%
  \BibitemOpen
  \bibfield{author}{%
  \bibinfo {author} {\bibfnamefont{A.}~\bibnamefont{Kovner}}\ and\ \bibinfo
  {author} {\bibfnamefont{M.}~\bibnamefont{Lublinsky}}}%
   (\bibinfo {year} {2012}),\
  \Eprint{http://arxiv.org/abs/1211.1928}{arXiv:1211.1928 [hep-ph]}%
  \bibAnnoteFile{NoStop}{Kovner:2012jm}%
\bibitem{Dusling:2012iga}%
  \BibitemOpen
  \bibfield{author}{%
  \bibinfo {author} {\bibfnamefont{K.}~\bibnamefont{Dusling}}\ and\ \bibinfo
  {author} {\bibfnamefont{R.}~\bibnamefont{Venugopalan}},\ }%
  \bibfield{journal}{%
  \Doi{10.1103/PhysRevLett.108.262001}{\bibinfo {journal} {Phys.Rev.Lett.}}\ }%
  \textbf{\bibinfo {volume} {108}},\ \bibinfo {pages} {262001} (\bibinfo {year}
  {2012}),\ \Eprint{http://arxiv.org/abs/1201.2658}{arXiv:1201.2658 [hep-ph]}%
  \bibAnnoteFile{NoStop}{Dusling:2012iga}%
\bibitem{Dumitru:2008wn}%
  \BibitemOpen
  \bibfield{author}{%
  \bibinfo {author} {\bibfnamefont{A.}~\bibnamefont{Dumitru}}, \bibinfo
  {author} {\bibfnamefont{F.}~\bibnamefont{Gelis}}, \bibinfo {author}
  {\bibfnamefont{L.}~\bibnamefont{McLerran}},\ and\ \bibinfo {author}
  {\bibfnamefont{R.}~\bibnamefont{Venugopalan}},\ }%
  \bibfield{journal}{%
  \Doi{10.1016/j.nuclphysa.2008.06.012}{\bibinfo {journal} {Nucl.Phys.}}\ }%
  \textbf{\bibinfo {volume} {A810}},\ \bibinfo {pages} {91} (\bibinfo {year}
  {2008}),\ \Eprint{http://arxiv.org/abs/0804.3858}{arXiv:0804.3858 [hep-ph]}%
  \bibAnnoteFile{NoStop}{Dumitru:2008wn}%
\bibitem{Dusling:2009ni}%
  \BibitemOpen
  \bibfield{author}{%
  \bibinfo {author} {\bibfnamefont{K.}~\bibnamefont{Dusling}}, \bibinfo
  {author} {\bibfnamefont{F.}~\bibnamefont{Gelis}}, \bibinfo {author}
  {\bibfnamefont{T.}~\bibnamefont{Lappi}},\ and\ \bibinfo {author}
  {\bibfnamefont{R.}~\bibnamefont{Venugopalan}},\ }%
  \bibfield{journal}{%
  \Doi{10.1016/j.nuclphysa.2009.12.044,
  10.1016/j.nuclphysa.2009.12.044}{\bibinfo {journal} {Nucl.Phys.}}\ }%
  \textbf{\bibinfo {volume} {A836}},\ \bibinfo {pages} {159} (\bibinfo {year}
  {2010}),\ \Eprint{http://arxiv.org/abs/0911.2720}{arXiv:0911.2720 [hep-ph]}%
  \bibAnnoteFile{NoStop}{Dusling:2009ni}%
\bibitem{Dumitru:2010iy}%
  \BibitemOpen
  \bibfield{author}{%
  \bibinfo {author} {\bibfnamefont{A.}~\bibnamefont{Dumitru}}, \bibinfo
  {author} {\bibfnamefont{K.}~\bibnamefont{Dusling}}, \bibinfo {author}
  {\bibfnamefont{F.}~\bibnamefont{Gelis}}, \bibinfo {author}
  {\bibfnamefont{J.}~\bibnamefont{Jalilian-Marian}}, \bibinfo {author}
  {\bibfnamefont{T.}~\bibnamefont{Lappi}}, \emph{et~al.},\ }%
  \bibfield{journal}{%
  \Doi{10.1016/j.physletb.2011.01.024}{\bibinfo {journal} {Phys.Lett.}}\ }%
  \textbf{\bibinfo {volume} {B697}},\ \bibinfo {pages} {21} (\bibinfo {year}
  {2011}),\ \Eprint{http://arxiv.org/abs/1009.5295}{arXiv:1009.5295 [hep-ph]}%
  \bibAnnoteFile{NoStop}{Dumitru:2010iy}%
\bibitem{Gelis:2010nm}%
  \BibitemOpen
  \bibfield{author}{%
  \bibinfo {author} {\bibfnamefont{F.}~\bibnamefont{Gelis}}, \bibinfo {author}
  {\bibfnamefont{E.}~\bibnamefont{Iancu}}, \bibinfo {author}
  {\bibfnamefont{J.}~\bibnamefont{Jalilian-Marian}},\ and\ \bibinfo {author}
  {\bibfnamefont{R.}~\bibnamefont{Venugopalan}},\ }%
  \bibfield{journal}{%
  \Doi{10.1146/annurev.nucl.010909.083629}{\bibinfo {journal}
  {Ann.Rev.Nucl.Part.Sci.}}\ }%
  \textbf{\bibinfo {volume} {60}},\ \bibinfo {pages} {463} (\bibinfo {year}
  {2010}),\ \Eprint{http://arxiv.org/abs/1002.0333}{arXiv:1002.0333 [hep-ph]}%
  \bibAnnoteFile{NoStop}{Gelis:2010nm}%
\bibitem{Dusling:2012cg}%
  \BibitemOpen
  \bibfield{author}{%
  \bibinfo {author} {\bibfnamefont{K.}~\bibnamefont{Dusling}}\ and\ \bibinfo
  {author} {\bibfnamefont{R.}~\bibnamefont{Venugopalan}}}%
   (\bibinfo {year} {2012}),\
  \Eprint{http://arxiv.org/abs/1210.3890}{arXiv:1210.3890 [hep-ph]}%
  \bibAnnoteFile{NoStop}{Dusling:2012cg}%
\bibitem{Gribov:1984tu}%
  \BibitemOpen
  \bibfield{author}{%
  \bibinfo {author} {\bibfnamefont{L.}~\bibnamefont{Gribov}}, \bibinfo {author}
  {\bibfnamefont{E.}~\bibnamefont{Levin}},\ and\ \bibinfo {author}
  {\bibfnamefont{M.}~\bibnamefont{Ryskin}},\ }%
  \bibfield{journal}{%
  \Doi{10.1016/0370-1573(83)90022-4}{\bibinfo {journal} {Phys.Rept.}}\ }%
  \textbf{\bibinfo {volume} {100}},\ \bibinfo {pages} {1} (\bibinfo {year}
  {1983})%
  \bibAnnoteFile{NoStop}{Gribov:1984tu}%
\bibitem{Mueller:1985wy}%
  \BibitemOpen
  \bibfield{author}{%
  \bibinfo {author} {\bibfnamefont{A.~H.}\ \bibnamefont{Mueller}}\ and\
  \bibinfo {author} {\bibfnamefont{J.-w.}\ \bibnamefont{Qiu}},\ }%
  \bibfield{journal}{%
  \Doi{10.1016/0550-3213(86)90164-1}{\bibinfo {journal} {Nucl.Phys.}}\ }%
  \textbf{\bibinfo {volume} {B268}},\ \bibinfo {pages} {427} (\bibinfo {year}
  {1986})%
  \bibAnnoteFile{NoStop}{Mueller:1985wy}%
\bibitem{Balitsky:1978ic}%
  \BibitemOpen
  \bibfield{author}{%
  \bibinfo {author} {\bibfnamefont{I.}~\bibnamefont{Balitsky}}\ and\ \bibinfo
  {author} {\bibfnamefont{L.}~\bibnamefont{Lipatov}},\ }%
  \bibfield{journal}{%
  \bibinfo {journal} {Sov.J.Nucl.Phys.}\ }%
  \textbf{\bibinfo {volume} {28}},\ \bibinfo {pages} {822} (\bibinfo {year}
  {1978})%
  \bibAnnoteFile{NoStop}{Balitsky:1978ic}%
\bibitem{Kuraev:1977fs}%
  \BibitemOpen
  \bibfield{author}{%
  \bibinfo {author} {\bibfnamefont{E.}~\bibnamefont{Kuraev}}, \bibinfo {author}
  {\bibfnamefont{L.}~\bibnamefont{Lipatov}},\ and\ \bibinfo {author}
  {\bibfnamefont{V.~S.}\ \bibnamefont{Fadin}},\ }%
  \bibfield{journal}{%
  \bibinfo {journal} {Sov.Phys.JETP}\ }%
  \textbf{\bibinfo {volume} {45}},\ \bibinfo {pages} {199} (\bibinfo {year}
  {1977})%
  \bibAnnoteFile{NoStop}{Kuraev:1977fs}%
\bibitem{CMS:2012qk}%
  \BibitemOpen
  \bibfield{author}{%
  \bibinfo {author} {\bibfnamefont{S.}~\bibnamefont{Chatrchyan}} \emph{et~al.}
  (\bibinfo {collaboration} {CMS Collaboration}),\ }%
  \bibfield{journal}{%
  \bibinfo {journal} {Physics Letters B}}%
   (\bibinfo {year} {2012}),\
  \Eprint{http://arxiv.org/abs/1210.5482}{arXiv:1210.5482 [nucl-ex]}%
  \bibAnnoteFile{NoStop}{CMS:2012qk}%
\bibitem{Note1}%
  \BibitemOpen
  \bibinfo {note} {Here onwards we will use the CMS notation $N_{\protect \rm
  trk}^{\protect \rm offline}$ to discuss the number of charged hadron tracks.
  Our results will be normalized to the same by equating our multiplicities to
  their values for the same in minimum bias proton-proton collisions. This
  point is discussed further on in the text.}%
  \bibAnnoteFile{Stop}{Note1}%
\bibitem{Note2}%
  \BibitemOpen
  \bibinfo {note} {As previously in~\cite {Dusling:2012cg}, the delta
  distribution is smeared as $\delta (\phi _{pq})\to \protect \frac
  {1}{\protect \sqrt { 2\pi \sigma }}e^{-\protect \frac {\phi _{pq}^2}{2\sigma
  ^2}}$, where $\Delta \phi _{p,q}=\phi _p-\phi _q$ and $\sigma =3\protect
  \textrm { GeV}/ {\protect \bf p}_T $ is a $ {\protect \bf p}_T $ dependent
  width on the order of the saturation scale. The associated yield--the
  integral over the near-side signal--is insensitive to details of this
  smearing.}%
  \bibAnnoteFile{Stop}{Note2}%
\bibitem{Note3}%
  \BibitemOpen
  \bibinfo {note} {Other parameters, which are held fixed in p+p and p+Pb are
  the transverse overlap area $S_\perp $ and the non-perturbative constant
  $\zeta =1/6$ that represents the effect of soft multigluon interactions, and
  is independently constrained by p+p multiplicity distributions~\cite
  {Tribedy:2010ab,Tribedy:2011aa} and real time classical Yang-Mills
  computations~\cite {Schenke:2012hg}.}%
  \bibAnnoteFile{Stop}{Note3}%
\bibitem{Balitsky:1995ub}%
  \BibitemOpen
  \bibfield{author}{%
  \bibinfo {author} {\bibfnamefont{I.}~\bibnamefont{Balitsky}},\ }%
  \bibfield{journal}{%
  \Doi{10.1016/0550-3213(95)00638-9}{\bibinfo {journal} {Nucl.Phys.}}\ }%
  \textbf{\bibinfo {volume} {B463}},\ \bibinfo {pages} {99} (\bibinfo {year}
  {1996}),\ \Eprint{http://arxiv.org/abs/hep-ph/9509348}{arXiv:hep-ph/9509348
  [hep-ph]}%
  \bibAnnoteFile{NoStop}{Balitsky:1995ub}%
\bibitem{Kovchegov:1999yj}%
  \BibitemOpen
  \bibfield{author}{%
  \bibinfo {author} {\bibfnamefont{Y.~V.}\ \bibnamefont{Kovchegov}},\ }%
  \bibfield{journal}{%
  \Doi{10.1103/PhysRevD.60.034008}{\bibinfo {journal} {Phys.Rev.}}\ }%
  \textbf{\bibinfo {volume} {D60}},\ \bibinfo {pages} {034008} (\bibinfo {year}
  {1999}),\ \Eprint{http://arxiv.org/abs/hep-ph/9901281}{arXiv:hep-ph/9901281
  [hep-ph]}%
  \bibAnnoteFile{NoStop}{Kovchegov:1999yj}%
\bibitem{Albacete:2010sy}%
  \BibitemOpen
  \bibfield{author}{%
  \bibinfo {author} {\bibfnamefont{J.~L.}\ \bibnamefont{Albacete}}, \bibinfo
  {author} {\bibfnamefont{N.}~\bibnamefont{Armesto}}, \bibinfo {author}
  {\bibfnamefont{J.~G.}\ \bibnamefont{Milhano}}, \bibinfo {author}
  {\bibfnamefont{P.}~\bibnamefont{Quiroga-Arias}},\ and\ \bibinfo {author}
  {\bibfnamefont{C.~A.}\ \bibnamefont{Salgado}},\ }%
  \bibfield{journal}{%
  \Doi{10.1140/epjc/s10052-011-1705-3}{\bibinfo {journal} {Eur.Phys.J.}}\ }%
  \textbf{\bibinfo {volume} {C71}},\ \bibinfo {pages} {1705} (\bibinfo {year}
  {2011}),\ \Eprint{http://arxiv.org/abs/1012.4408}{arXiv:1012.4408 [hep-ph]}%
  \bibAnnoteFile{NoStop}{Albacete:2010sy}%
\bibitem{Colferai:2010wu}%
  \BibitemOpen
  \bibfield{author}{%
  \bibinfo {author} {\bibfnamefont{D.}~\bibnamefont{Colferai}}, \bibinfo
  {author} {\bibfnamefont{F.}~\bibnamefont{Schwennsen}}, \bibinfo {author}
  {\bibfnamefont{L.}~\bibnamefont{Szymanowski}},\ and\ \bibinfo {author}
  {\bibfnamefont{S.}~\bibnamefont{Wallon}},\ }%
  \bibfield{journal}{%
  \Doi{10.1007/JHEP12(2010)026}{\bibinfo {journal} {JHEP}}\ }%
  \textbf{\bibinfo {volume} {1012}},\ \bibinfo {pages} {026} (\bibinfo {year}
  {2010}),\ \Eprint{http://arxiv.org/abs/1002.1365}{arXiv:1002.1365 [hep-ph]}%
  \bibAnnoteFile{NoStop}{Colferai:2010wu}%
\bibitem{Fadin:1996zv}%
  \BibitemOpen
  \bibfield{author}{%
  \bibinfo {author} {\bibfnamefont{V.~S.}\ \bibnamefont{Fadin}}, \bibinfo
  {author} {\bibfnamefont{M.}~\bibnamefont{Kotsky}},\ and\ \bibinfo {author}
  {\bibfnamefont{L.}~\bibnamefont{Lipatov}}}%
   (\bibinfo {year} {1996}),\
  \Eprint{http://arxiv.org/abs/hep-ph/9704267}{arXiv:hep-ph/9704267 [hep-ph]}%
  \bibAnnoteFile{NoStop}{Fadin:1996zv}%
\bibitem{Kniehl:2000fe}%
  \BibitemOpen
  \bibfield{author}{%
  \bibinfo {author} {\bibfnamefont{B.~A.}\ \bibnamefont{Kniehl}}, \bibinfo
  {author} {\bibfnamefont{G.}~\bibnamefont{Kramer}},\ and\ \bibinfo {author}
  {\bibfnamefont{B.}~\bibnamefont{Potter}},\ }%
  \bibfield{journal}{%
  \Doi{10.1016/S0550-3213(00)00303-5}{\bibinfo {journal} {Nucl. Phys.}}\ }%
  \textbf{\bibinfo {volume} {B582}},\ \bibinfo {pages} {514} (\bibinfo {year}
  {2000}),\ \Eprint{http://arxiv.org/abs/hep-ph/0010289}{arXiv:hep-ph/0010289}%
  \bibAnnoteFile{NoStop}{Kniehl:2000fe}%
\bibitem{Note4}%
  \BibitemOpen
  \bibinfo {note} {Replacing the rapidity $y$ with the pseudo-rapidity $\eta $
  is a good approximation for the $p_T$, $q_T$ of interest.}%
  \bibAnnoteFile{Stop}{Note4}%
\bibitem{Bozek:2011if}%
  \BibitemOpen
  \bibfield{author}{%
  \bibinfo {author} {\bibfnamefont{P.}~\bibnamefont{Bozek}},\ }%
  \bibfield{journal}{%
  \Doi{10.1103/PhysRevC.85.014911}{\bibinfo {journal} {Phys.Rev.}}\ }%
  \textbf{\bibinfo {volume} {C85}},\ \bibinfo {pages} {014911} (\bibinfo {year}
  {2012}),\ \Eprint{http://arxiv.org/abs/1112.0915}{arXiv:1112.0915 [hep-ph]}%
  \bibAnnoteFile{NoStop}{Bozek:2011if}%
\bibitem{Albacete:2012xq}%
  \BibitemOpen
  \bibfield{author}{%
  \bibinfo {author} {\bibfnamefont{J.~L.}\ \bibnamefont{Albacete}}, \bibinfo
  {author} {\bibfnamefont{A.}~\bibnamefont{Dumitru}}, \bibinfo {author}
  {\bibfnamefont{H.}~\bibnamefont{Fujii}},\ and\ \bibinfo {author}
  {\bibfnamefont{Y.}~\bibnamefont{Nara}}}%
   (\bibinfo {year} {2012}),\
  \Eprint{http://arxiv.org/abs/1209.2001}{arXiv:1209.2001 [hep-ph]}%
  \bibAnnoteFile{NoStop}{Albacete:2012xq}%
\bibitem{Dumitru:2001ux}%
  \BibitemOpen
  \bibfield{author}{%
  \bibinfo {author} {\bibfnamefont{A.}~\bibnamefont{Dumitru}}\ and\ \bibinfo
  {author} {\bibfnamefont{L.~D.}\ \bibnamefont{McLerran}},\ }%
  \bibfield{journal}{%
  \Doi{10.1016/S0375-9474(01)01301-X}{\bibinfo {journal} {Nucl.Phys.}}\ }%
  \textbf{\bibinfo {volume} {A700}},\ \bibinfo {pages} {492} (\bibinfo {year}
  {2002}),\ \Eprint{http://arxiv.org/abs/hep-ph/0105268}{arXiv:hep-ph/0105268
  [hep-ph]}%
  \bibAnnoteFile{NoStop}{Dumitru:2001ux}%
\bibitem{Note5}%
  \BibitemOpen
  \bibinfo {note} {We have checked that this scaling is approximately satisfied
  by our computed single inclusive multiplicity.}%
  \bibAnnoteFile{Stop}{Note5}%
\bibitem{Dumitru:2010mv}%
  \BibitemOpen
  \bibfield{author}{%
  \bibinfo {author} {\bibfnamefont{A.}~\bibnamefont{Dumitru}}\ and\ \bibinfo
  {author} {\bibfnamefont{J.}~\bibnamefont{Jalilian-Marian}},\ }%
  \bibfield{journal}{%
  \Doi{10.1103/PhysRevD.81.094015}{\bibinfo {journal} {Phys.Rev.}}\ }%
  \textbf{\bibinfo {volume} {D81}},\ \bibinfo {pages} {094015} (\bibinfo {year}
  {2010}),\ \Eprint{http://arxiv.org/abs/1001.4820}{arXiv:1001.4820 [hep-ph]}%
  \bibAnnoteFile{NoStop}{Dumitru:2010mv}%
\bibitem{Dumitru:2011zz}%
  \BibitemOpen
  \bibfield{author}{%
  \bibinfo {author} {\bibfnamefont{A.}~\bibnamefont{Dumitru}}, \bibinfo
  {author} {\bibfnamefont{J.}~\bibnamefont{Jalilian-Marian}},\ and\ \bibinfo
  {author} {\bibfnamefont{E.}~\bibnamefont{Petreska}},\ }%
  \bibfield{journal}{%
  \Doi{10.1103/PhysRevD.84.014018}{\bibinfo {journal} {Phys.Rev.}}\ }%
  \textbf{\bibinfo {volume} {D84}},\ \bibinfo {pages} {014018} (\bibinfo {year}
  {2011}),\ \Eprint{http://arxiv.org/abs/1105.4155}{arXiv:1105.4155 [hep-ph]}%
  \bibAnnoteFile{NoStop}{Dumitru:2011zz}%
\bibitem{Kovner:2010xk}%
  \BibitemOpen
  \bibfield{author}{%
  \bibinfo {author} {\bibfnamefont{A.}~\bibnamefont{Kovner}}\ and\ \bibinfo
  {author} {\bibfnamefont{M.}~\bibnamefont{Lublinsky}},\ }%
  \bibfield{journal}{%
  \Doi{10.1103/PhysRevD.83.034017}{\bibinfo {journal} {Phys.Rev.}}\ }%
  \textbf{\bibinfo {volume} {D83}},\ \bibinfo {pages} {034017} (\bibinfo {year}
  {2011}),\ \Eprint{http://arxiv.org/abs/1012.3398}{arXiv:1012.3398 [hep-ph]}%
  \bibAnnoteFile{NoStop}{Kovner:2010xk}%
\bibitem{Kovner:2011pe}%
  \BibitemOpen
  \bibfield{author}{%
  \bibinfo {author} {\bibfnamefont{A.}~\bibnamefont{Kovner}}\ and\ \bibinfo
  {author} {\bibfnamefont{M.}~\bibnamefont{Lublinsky}}}%
   (\bibinfo {year} {2011}),\
  \Eprint{http://arxiv.org/abs/1109.0347}{arXiv:1109.0347 [hep-ph]}%
  \bibAnnoteFile{NoStop}{Kovner:2011pe}%
\bibitem{Note6}%
  \BibitemOpen
  \bibinfo {note} {Data is also available for $p_T < 1$ GeV; we have not
  compared the theory predictions to data in this window because the theory
  systematic errors are large.}%
  \bibAnnoteFile{Stop}{Note6}%
\bibitem{Note7}%
  \BibitemOpen
  \bibinfo {note} {There is a genuine subtlety in comparison of the glasma
  graph $K$-factors in p+p and p+Pb. In Fig.~(2) of \cite {Dusling:2012cg}, the
  BFKL contribution on the {\protect \it nearside} is ZYAM'ed out and the
  glasma contribution alone with $K=1$ gives a good fit to data. In Fig.~(4),
  for the matrix, as discussed in the text of \cite {Dusling:2012cg}, the BFKL
  contribution gives a tiny (relative to awayside) nearside
  {``anti-collimation"} which is compensated by cranking up the glasma $K$
  factor to 2.3. However, the BFKL calculation is not reliable on the nearside
  and further, it contaminates the tiny but unmistakable glasma signal. If we
  make it flat from 0 to $\Delta \phi \sim \pi /2$, all the systematics of p+p
  (as in Fig. 2 of \cite {Dusling:2012cg}) would be reproduced with a glasma
  $K$ factor of unity. In the p+Pb case, because the glasma signal is so large
  for high multiplicity windows, the uncertainties in BFKL on the nearside are
  of not much import and glasma $K=1$ works quite well.}%
  \bibAnnoteFile{Stop}{Note7}%
\bibitem{Caporale:2012qd}%
  \BibitemOpen
  \bibfield{author}{%
  \bibinfo {author} {\bibfnamefont{F.}~\bibnamefont{Caporale}}, \bibinfo
  {author} {\bibfnamefont{D.~Y.}\ \bibnamefont{Ivanov}}, \bibinfo {author}
  {\bibfnamefont{B.}~\bibnamefont{Murdaca}},\ and\ \bibinfo {author}
  {\bibfnamefont{A.}~\bibnamefont{Papa}}}%
   (\bibinfo {year} {2012}),\
  \Eprint{http://arxiv.org/abs/1209.6233}{arXiv:1209.6233 [hep-ph]}%
  \bibAnnoteFile{NoStop}{Caporale:2012qd}%
\bibitem{Note8}%
  \BibitemOpen
  \bibinfo {note} {This may be contrasted to the collimated NLLx contributions
  that are 10\%-30\% corrections to LLx, and will be significantly accounted
  for already by the running coupling effects we have included.}%
  \bibAnnoteFile{Stop}{Note8}%
\bibitem{Kowalski:2006hc}%
  \BibitemOpen
  \bibfield{author}{%
  \bibinfo {author} {\bibfnamefont{H.}~\bibnamefont{Kowalski}}, \bibinfo
  {author} {\bibfnamefont{L.}~\bibnamefont{Motyka}},\ and\ \bibinfo {author}
  {\bibfnamefont{G.}~\bibnamefont{Watt}},\ }%
  \bibfield{journal}{%
  \Doi{10.1103/PhysRevD.74.074016}{\bibinfo {journal} {Phys.Rev.}}\ }%
  \textbf{\bibinfo {volume} {D74}},\ \bibinfo {pages} {074016} (\bibinfo {year}
  {2006}),\ \Eprint{http://arxiv.org/abs/hep-ph/0606272}{arXiv:hep-ph/0606272
  [hep-ph]}%
  \bibAnnoteFile{NoStop}{Kowalski:2006hc}%
\bibitem{Tribedy:2010ab}%
  \BibitemOpen
  \bibfield{author}{%
  \bibinfo {author} {\bibfnamefont{P.}~\bibnamefont{Tribedy}}\ and\ \bibinfo
  {author} {\bibfnamefont{R.}~\bibnamefont{Venugopalan}},\ }%
  \bibfield{journal}{%
  \Doi{10.1016/j.nuclphysa.2010.12.006,
  10.1016/j.nuclphysa.2011.04.008}{\bibinfo {journal} {Nucl.Phys.}}\ }%
  \textbf{\bibinfo {volume} {A850}},\ \bibinfo {pages} {136} (\bibinfo {year}
  {2011}),\ \Eprint{http://arxiv.org/abs/1011.1895}{arXiv:1011.1895 [hep-ph]}%
  \bibAnnoteFile{NoStop}{Tribedy:2010ab}%
\bibitem{Note9}%
  \BibitemOpen
  \bibinfo {note} {For the qualitative arguments in this section, this
  assumption is sufficient because the glasma signal, to first approximation,
  is boost invariant.}%
  \bibAnnoteFile{Stop}{Note9}%
\bibitem{Note10}%
  \BibitemOpen
  \bibinfo {note} {Define $f( {\protect \bf k}_T )=\Phi _A( {\protect \bf k}_T
  )\Phi _B(\protect \vert {\protect \bf p}_T + {\protect \bf k}_T \protect
  \vert )$ and $g( {\protect \bf k}_T )=\Phi _A( {\protect \bf k}_T )\Phi
  _B(\protect \vert {\protect \bf q}_T + {\protect \bf k}_T \protect \vert )$.
  The two dimensional form of the Cauchy–-Schwarz inequality is \begin
  {align} \DOTSI \intop \ilimits@ d^2k_\perp \protect \tmspace +\thickmuskip
  {.2777em}f( {\protect \bf k}_T )g( {\protect \bf k}_T )\leq \protect \sqrt
  {\DOTSI \intop \ilimits@ d^2 {\protect \bf k}_T \protect \tmspace
  +\thickmuskip {.2777em}\left |f( {\protect \bf k}_T )\right |^2} \protect
  \sqrt {\DOTSI \intop \ilimits@ d^2 {\protect \bf k}_T \protect \tmspace
  +\thickmuskip {.2777em}\left |g( {\protect \bf k}_T )\right |^2}\protect
  \nonumber \end {align} with the equality holding iff $f( {\protect \bf k}_T
  )\propto g( {\protect \bf k}_T )$.}%
  \bibAnnoteFile{Stop}{Note10}%
\bibitem{Note11}%
  \BibitemOpen
  \bibinfo {note} {This is confirmed by a simple analysis treating the $\Phi
  $'s as Gaussian wavefunctions.}%
  \bibAnnoteFile{Stop}{Note11}%
\bibitem{Note12}%
  \BibitemOpen
  \bibinfo {note} {In addition to Eq.~(\ref {eq:discuss1}), there is a
  contribution that has a maximum at $\pi $ and a minimum at $0$. Its the sum
  of the two contributions that gives the {``dipole"}-like signal with a
  minimum at $\pi /2$.}%
  \bibAnnoteFile{Stop}{Note12}%
\bibitem{Note13}%
  \BibitemOpen
  \bibinfo {note} {Jacobian factors cancel between numerator and denominator.}%
  \bibAnnoteFile{Stop}{Note13}%
\bibitem{Note14}%
  \BibitemOpen
  \bibinfo {note} {The square root in the denominator implies finite support
  for $Q_B$ in the narrow range $1 \leq Q_B/\protect \sqrt {2p_T^2+2Q_A^2} \leq
  \protect \sqrt {2}$, which is an artifact of taking Delta function
  approximation for the wave-functions.}%
  \bibAnnoteFile{Stop}{Note14}%
\bibitem{Gelis:2008sz}%
  \BibitemOpen
  \bibfield{author}{%
  \bibinfo {author} {\bibfnamefont{F.}~\bibnamefont{Gelis}}, \bibinfo {author}
  {\bibfnamefont{T.}~\bibnamefont{Lappi}},\ and\ \bibinfo {author}
  {\bibfnamefont{R.}~\bibnamefont{Venugopalan}},\ }%
  \bibfield{journal}{%
  \Doi{10.1103/PhysRevD.79.094017}{\bibinfo {journal} {Phys.Rev.}}\ }%
  \textbf{\bibinfo {volume} {D79}},\ \bibinfo {pages} {094017} (\bibinfo {year}
  {2009}),\ \Eprint{http://arxiv.org/abs/0810.4829}{arXiv:0810.4829 [hep-ph]}%
  \bibAnnoteFile{NoStop}{Gelis:2008sz}%
\bibitem{Bozek:2012gr}%
  \BibitemOpen
  \bibfield{author}{%
  \bibinfo {author} {\bibfnamefont{P.}~\bibnamefont{Bozek}}\ and\ \bibinfo
  {author} {\bibfnamefont{W.}~\bibnamefont{Broniowski}}}%
   (\bibinfo {year} {2012}),\
  \Eprint{http://arxiv.org/abs/1211.0845}{arXiv:1211.0845 [nucl-th]}%
  \bibAnnoteFile{NoStop}{Bozek:2012gr}%
\bibitem{Tribedy:2011aa}%
  \BibitemOpen
  \bibfield{author}{%
  \bibinfo {author} {\bibfnamefont{P.}~\bibnamefont{Tribedy}}\ and\ \bibinfo
  {author} {\bibfnamefont{R.}~\bibnamefont{Venugopalan}}}%
   (\bibinfo {year} {2011}),\
  \Eprint{http://arxiv.org/abs/1112.2445}{arXiv:1112.2445 [hep-ph]}%
  \bibAnnoteFile{NoStop}{Tribedy:2011aa}%
\bibitem{Schenke:2012hg}%
  \BibitemOpen
  \bibfield{author}{%
  \bibinfo {author} {\bibfnamefont{B.}~\bibnamefont{Schenke}}, \bibinfo
  {author} {\bibfnamefont{P.}~\bibnamefont{Tribedy}},\ and\ \bibinfo {author}
  {\bibfnamefont{R.}~\bibnamefont{Venugopalan}}}%
   (\bibinfo {year} {2012}),\
  \Eprint{http://arxiv.org/abs/1206.6805}{arXiv:1206.6805 [hep-ph]}%
  \bibAnnoteFile{NoStop}{Schenke:2012hg}%
\end{thebibliography}
\end{document}